\documentclass{iopart}
\usepackage{iopams}
\usepackage{epsfig}
\usepackage{graphicx}
\newcommand{\rmeh}{{\rm e}}
\newcommand{\rmih}{{\rm i}}

\newcommand{\bra}[1]{\langle#1|}
\newcommand{\ket}[1]{|#1\rangle}
\newcommand{\beq}{\begin{equation}}
\newcommand{\eeq}{\end{equation}}

\newcommand{\fat}[1]{\mbox{\boldmath$#1$\unboldmath}}         
\newcommand{\ham}{\hat{H}}                          
\newcommand{\imag}{{\rm i}}                        
\newcommand{\dm}{\hat{\rho}}                        
\newcommand{\bcr}[1]{b^{\dagger}_{#1}}                        
\newcommand{\ban}[1]{b^{\phantom{\dagger}}_{#1}}              

\newcommand{\ent}{\mathcal S}

\begin{document}
\title[Adaptive time-dependent DMRG]{Time-dependent density-matrix renormalization-group using 
adaptive effective Hilbert spaces}
\author{A J Daley\dag, C Kollath\ddag, U. Schollw\"{o}ck\S\ and G Vidal$\|$}
\address{\dag\ Institut f\"{u}r Theoretische Physik, Universit\"{a}t Innsbruck, and \\ Institute for Quantum Optics and Quantum Information of the Austrian Academy of Sciences, 
A-6020 Innsbruck, Austria}
\address{\ddag\ Department f\"{u}r Physik, Ludwig-Maximilians-Universit\"{a}t M\"{u}nchen, D-80333
Munich, Germany}
\address{\S\ Institut f\"{u}r Theoretische Physik C, RWTH Aachen, D-52056 Aachen, Germany}
\address{$\|$ Institute for Quantum Information, California Institute 
of Technology, Pasadena, CA 91125, USA}
\date{\today}

\begin{abstract}
An algorithm for the simulation of the evolution of slightly entangled 
quantum states has been recently proposed as a tool to study 
time-dependent phenomena in one-dimensional quantum systems. 
Its key feature is a time-evolving block-decimation (TEBD) procedure 
to identify and dynamically update the relevant, conveniently small 
subregion of the otherwise exponentially large Hilbert space. Potential 
applications of the TEBD algorithm are the simulation of time-dependent 
Hamiltonians, transport in quantum systems far from equilibrium and 
dissipative quantum mechanics. In this paper we translate the TEBD 
algorithm into the language of matrix product states in order to both 
highlight and exploit its resemblances to the widely used density-matrix 
renormalization-group (DMRG) algorithms. The TEBD algorithm, being based 
on updating a matrix product state in time, is very accessible to the 
DMRG community and it can be enhanced by using well-known DMRG techniques, 
for instance in the event of good quantum numbers. More importantly, we 
show how it can be simply incorporated into existing DMRG implementations 
to produce a remarkably effective and versatile ``adaptive time-dependent 
DMRG'' variant, that we also test and compare to previous proposals.
\end{abstract}
\submitto{JSTAT}
\maketitle

\section{Introduction}
\label{sec:intro}
Over many decades the description of the physical properties of 
low-dimensional strongly correlated quantum systems has been one of 
the major tasks in theoretical condensed matter physics. Generically, 
this task is complicated by the strong quantum fluctuations present in 
such systems which are 
usually modelled by minimal-model Hubbard or Heisenberg-style 
Hamiltonians. Despite the apparent simplicity of these Hamiltonians, 
few analytically exact solutions are available and most analytical 
approximations remain uncontrolled. Hence, numerical approaches have 
always been of particular interest, among them exact diagonalization 
and quantum Monte Carlo.

Decisive progress in the description of the low-energy equilibrium 
properties of one-dimensional strongly correlated quantum Hamiltonians 
was achieved by the invention of the density-matrix 
renormalization-group (DMRG) \cite{Whit92,Whit93}. It is concerned 
with the iterative decimation of the Hilbert space of a growing 
quantum system such that some quantum state, say the ground state, is 
approximated in that restricted space with a maximum of overlap with the true 
state. Let the quantum state of a one-dimensional system be 
\begin{equation}
\ket{\psi}= \sum_{i} \sum_{j} \psi_{ij} 
\ket{i}\ket{j},
\label{eq:groundsimple}
\end{equation}
where we consider a partition of the system into two blocks S and E, and where $\{\ket{i}\}$ 
and $\{\ket{j}\}$ are orthonormal bases of S and E respectively. Then
the DMRG decimation procedure consists of projecting $\ket{\psi}$ on the 
Hilbert spaces for S and E spanned by the $M$ eigenvectors 
$\ket{w_{\alpha}^S}$ and  $\ket{w_{\alpha}^E}$
corresponding to the largest eigenvalues $\lambda_{\alpha}^2$ 
of the reduced density matrices
\begin{equation}
    \dm_{S} = {\rm Tr}_{E} \ket{\psi}\bra{\psi} \quad\quad 
    \dm_{E} = {\rm Tr}_{S} \ket{\psi}\bra{\psi} ,     
\end{equation}
such that $\dm_{S} \ket{w_{\alpha}^S} = \lambda_{\alpha}^2 \ket{w_{\alpha}^S}$
and $\dm_{E} \ket{w_{\alpha}^E} = \lambda_{\alpha}^2 
\ket{w_{\alpha}^E}$. That both density matrices have the same 
eigenvalue spectrum is reflected in the guaranteed existence of the 
so-called Schmidt decomposition of the wave function \cite{QC00}, 
\begin{equation}
    \ket{\psi} = \sum_{\alpha} \lambda_{\alpha} 
    \ket{w_{\alpha}^S}\ket{w_{\alpha}^E} , \quad \lambda_{\alpha}\geq 
    0 ,
\label{eq:Schmidt}
\end{equation}
where the number of positive $\lambda_{\alpha}$ is bounded by the 
dimension of the smaller of the bases of S and E.

Recently \cite{Osbo02a,Daws04,Hase03,Vida03c,Gait03,Lato03}, the ability of the DMRG decimation procedure to preserve 
the entanglement of $\ket{\psi}$ between S and E has been studied in the 
context of quantum information science \cite{QC00,Benn00}. This blooming 
field of research, bridging between quantum physics, computer science and 
information theory, offers a novel conceptual framework for the study of quantum 
many-body systems 
\cite{QC00,Osbo02a,Daws04,Hase03,Vida03c,Gait03,Lato03,Benn00,Osbo02,Oste02,Lege03,Lege04,Vers04,Vers04a,Vida03b}. 
New insights into old quantum many-body problems can be gained from the 
perspective of quantum information science, mainly through its eagerness 
to characterize quantum 
correlations. As an example, a better 
understanding of the reasons of the breakdown of the DMRG in two-dimensional
systems has been obtained in terms of the growth of bipartite entanglement 
in such systems \cite{Vida03c,Lato03}. 

More specifically, in quantum information the entanglement of 
$\ket{\psi}$ between S and E is quantified by the von Neumann entropy of $\dm_S$ 
(equivalently, of $\dm_E$),
\begin{equation}
\ent(\dm_S) = -\sum \lambda_\alpha^2 \log_2 \lambda_\alpha^2, 
\end{equation}
a quantity that imposes a useful (information theoretical) bound $M \geq 2^{\ent}$
on the minimal number $M$ of states to be kept during the DMRG decimation process 
if the truncated state is to be similar to $\ket{\psi}$. On the other hand, arguments 
from field theory imply that, at zero temperature, 
strongly correlated quantum systems are in some sense only 
slightly entangled in $d=1$ dimensions but significantly more entangled in 
$d>1$ dimensions: In particular, in $d=1$ a block corresponding to $l$ sites of a 
gapped infinite-length chain has an entropy $\ent_l$ that stays finite even in the 
thermodynamical limit $l\rightarrow \infty$, while at criticality $\ent_l$ only 
grows logarithmically with $l$. It is this saturation or, at most, 
moderate growth of $\ent_l$ that ultimately accounts for the succeess of DMRG 
in $d=1$. Instead, in the general $d$-dimensional 
case the entropy of bipartite entanglement for a block of linear 
dimension $l$ scales as $\ent_l \sim l^{d-1}$. Thus, in $d = 2$ dimensions the 
DMRG algorithm should keep a 
number $M$ of states that grows exponentially with $l$, and the simulation 
becomes inefficient for large $l$ (while still feasible for small $l$).

While DMRG 
has yielded an enormous wealth of information on the static and 
dynamic equilibrium properties of one-dimensional 
systems\cite{Pesc99,Scho04} and is arguably the most powerful method 
in the field, only few attempts have been 
made so far to determine the time evolution of the states of such systems, 
notably in a seminal paper by Cazalilla and Marston \cite{Caza02}.
This question is of relevance in the context of the time-dependent 
Hamiltonians realized e.g.\ in cold atoms in optical lattices 
\cite{Grei02, Stoef03}, in systems far from equilibrium in quantum transport,
or in dissipative quantum mechanics. 
However, in another example of how quantum information science can contribute 
to the study of quantum many-body physics, one of us (G.V.) has recently 
developed an algorithm for the simulation of slightly entangled quantum 
computations \cite{Vida03a} that can be used to simulate time evolutions of 
one-dimensional systems \cite{Vida03b}. 

This new algorithm, henceforth referred to as the time-evolving block 
decimation (TEBD) algorithm, considers a small, dynamically updated 
subspace of the blocks S and E in Eq. (\ref{eq:Schmidt}) to 
efficiently represent the state of the system, as we will review 
in detail below. It was originally developed in order to show that
a large amount of entanglement is necessary in quantum computations, 
the rationale there being quite simple: any {\em quantum} evolution 
(e.g. a quantum computation) involving only a ``sufficiently restricted'' 
amount of entanglement can be efficiently simulated in a {\em classical} 
computer using the TEBD algorithm; therefore, from an algorithmical point 
of view, any such quantum evolution is not more powerful than a classical 
computation. 

Regardless of the implications for computer science, the above connection 
between the amount of entanglement and the complexity of simulating quantum 
systems is of obvious practical interest in condensed matter physics since, 
for instance, in $d=1$ dimensions the entanglement of most quantum systems 
happens to be ``sufficiently restricted'' precisely in the sense required 
for the TEBD algorithm to yield an efficient simulation. In particular, the 
algorithm has already been implemented and tested successfully on spin chains\cite{Vida03b}, 
the Bose-Hubbard model and single-atom transistors\cite{JaZo04} and 
dissipative systems at finite temperature \cite{Zwo04}.

A primary aim of this paper is to reexpress the TEBD algorithm in a 
language more familiar to the DMRG community than the one originally 
used in Refs. \cite{Vida03b,Vida03a}, which made substantial use of the 
quantum information parlance. This turns out to be a rewarding task since, 
as we show, the conceptual and formal similarities between the TEBD and 
DMRG are extensive. Both algorithms search for an approximation to the 
true wave function within a restricted class of wave functions, which can 
be identified as matrix product states \cite{Klum93},
and had also been previously proposed under the name of 
finitely-correlated states\cite{Fann92}. Arguably, the big 
advantage of the TEBD algorithm relies on its flexibility to 
flow in time through the submanifold of matrix product states. 
Instead of considering time evolutions within some restricted 
subspace according to a fixed, projected, effective Hamiltonian, the 
TEBD algorithm updates a matrix product state in time using the
bare Hamiltonian directly. Thus, in a sense, it is the Schr\"odinger 
equation that decides, at each time step, which are the relevant 
eigenvectors for S and E in Eq. (\ref{eq:Schmidt}), as opposed 
to having to select them from some relatively small, pre-selected 
subspace.

A second goal of this paper is to show how the two algorithms can be 
integrated. The TEBD algorithm can be improved by considering 
well-known DMRG techniques, such as the handling of good quantum numbers. 
But most importantly, we will describe how the TEBD simulation algorithm 
can be incorporated into preexisting, quite widely used DMRG implementations, 
the so-called finite-system algorithm\cite{Whit93} using 
White's prediction algorithm\cite{Whit96}. The net result 
is an extremely powerful ``adaptive time-dependent DMRG'' algorithm, 
that we test and compare against previous proposals.

The outline of this paper is as follows: In Section \ref{sec:oldtime},
we discuss the problems currently encountered in applying DMRG to the 
calculation of explicitly time-dependent quantum states. Section
\ref{sec:matrixstates} reviews the common language of matrix product 
states. We then express both the TEBD simulation algorithm (Sec.\ 
\ref{sec:vidal}) and DMRG (Sec.\ \ref{sec:dmrg}) in this 
language, revealing where both methods coincide, where they differ 
and how they can be combined. In Section \ref{sec:newtime}, we then 
formulate the modifications to introduce the TEBD algorithm into 
standard DMRG to obtain the
adaptive time-dependent DMRG, and Section \ref{sec:bose} 
discusses an example application, concerning the quantum phase 
transition between a superfluid and a Mott-insulating state in
a Bose-Hubbard model. To conclude, we discuss in Section 
\ref{sec:conclusion} the potential of the new DMRG variant.

\section{Simulation of time-dependent quantum phenomena using DMRG}
\label{sec:oldtime}
The first attempt to simulate the time evolution of quantum states 
using DMRG is due to Cazalilla and Marston \cite{Caza02}. 
After applying a standard DMRG calculation using the Hamiltonian 
$\ham(t=0)$ to obtain the ground state of the system at $t=0$,
$\ket{\psi_{0}}$, the time-dependent Schr\"{o}dinger equation is 
numerically integrated forward in time, building an effective 
$\ham_{{\rm eff}}(t)=\ham_{{\rm eff}}(0)+\hat{V}_{{\rm eff}}(t)$, where 
$\ham_{{\rm eff}}(0)$ is 
taken as the Hamiltonian approximating
$\ham(0)$ in the truncated Hilbert space generated by DMRG. 
$\hat{V}_{{\rm eff}}(t)$ as an approximation to 
$\hat{V}(t)$ is built using 
the representations of operators in the block bases obtained in the 
standard DMRG calculation of the $t=0$ state. $\hat{V}(t)$ 
contains the changes in the Hamiltonian with respect to the starting
Hamiltonian: $\hat{H}(t)=\hat{H}_0+\hat{V}(t)$. The (effective) 
time-dependent Schr\"{o}dinger equation reads
\begin{equation}
    i\frac{\partial}{\partial t} \ket{\psi(t)} =
    [\ham_{{\rm eff}}-E_{0}+\hat{V}_{{\rm eff}}(t)] \ket{\psi(t)} ,
    \label{eq:Schroedinger}
\end{equation}
where the time-dependence of the ground state resulting of $\ham(0)$ has been
transformed away. If the evolution of the ground state is looked for, 
the initial condition is obviously to take 
$\ket{\psi(0)}=\ket{\psi_{0}}$ obtained by the preliminary DMRG run.
Forward integration can be carried out by step-size adaptive methods 
such as the Runge-Kutta integration based on the infinitesimal 
time evolution operator
\begin{equation}
    \ket{\psi(t+\delta t)} = (1- \imag 
    \ham(t) \delta t) \ket{\psi(t)},     
    \label{eq:nonunitaryevolution}
\end{equation}
where we drop the subscript denoting that we are dealing with 
effective Hamiltonians only. The algorithm used was a fourth-order adaptive
size Runge-Kutta algorithm \cite{recipes}.

Sources of errors in this approach are twofold, due to the 
approximations involved in numerically carrying out the time 
evolution, and to the fact that all operators live on a truncated 
Hilbert space.

For the systems studied we have obtained 
a conceptually simple improvement concerning the time
evolution by replacing the 
explicitly non-unitary time-evolution of Eq.\ 
(\ref{eq:nonunitaryevolution}) by the unitary Crank-Nicholson time 
evolution
\begin{equation}
    \ket{\psi(t+\delta t)} = \frac{1- \imag 
    \ham(t) \delta t/2}{1+ \imag \ham(t) \delta t/2} \ket{\psi(t)}.     
    \label{eq:unitaryevolution}
\end{equation}

To implement the Crank-Nicholson time evolution efficiently 
we have used a (non-Hermitian) biconjugate gradient method to 
calculate the denominator of Eq.\ (\ref{eq:unitaryevolution}).
In fact, 
this modification ensures higher precision of 
correlators, and the occurence of asymmetries with respect to reflection in the
results decreased. 

It should be noted, however, 
that for the Crank-Nicholson approach only lowest-order 
expansions of the time evolution operator $\exp (-\imag \ham \delta t)$ have
been taken; we have not pursued feasible higher-order expansions.

As a testbed for time-dependent DMRG methods we use throughout this 
paper the time-dependent Bose-Hubbard Hamiltonian,
\begin{equation}
    \ham_{BH}(t) = -J \sum_{i=1}^{L-1}  \bcr{i+1}\ban{i} + \bcr{i}\ban{i+1} + 
    \frac{U(t)}{2} \sum_{i=1}^L n_{i}(n_{i}-1),
    \label{eq:bosehubbard}
\end{equation}
where the (repulsive) onsite interaction $U>0$ is taken to be 
time-dependent. This model exhibits for commensurate filling
a Kosterlitz-Thouless-like 
quantum phase transition from a superfluid phase for $u<u_{c}$ (with $u=U/J$)
to a Mott-insulating phase for $u>u_{c}$. We have studied a Bose-Hubbard model with 
$L=8$ and open boundary conditions, 
total particle number $N=8$, $J=1$, and instantaneous switching from $U_{1}=2$ in 
the superfluid phase to $U_{2}=40$ in the Mott phase at $t=0$. We 
consider the nearest-neighbor correlation, a robust numerical quantity,
between sites 2 and 3. Up to 8 bosons per site (i.e.\ 
$N_{{\rm site}}=9$ states per site) were allowed to avoid cut-off 
effects in the bosonic occupation number in all calculations in this 
Section. All times in this paper are 
measured in units of $\hbar/J$ or $1/J$, setting $\hbar\equiv 1$.
Comparing Runge-Kutta and 
Crank-Nicholson (with time steps of $\delta t= 5 \times 10^{-5}$) we 
found the latter to be numerically preferable; all static 
time-dependent DMRG calculations have been carried out using the 
latter approach. 

However, Hilbert space truncation is at the origin of more severe 
approximations.
The key assumption underlying the approach of Cazalilla and Marston is 
that the effective static Hilbert space created in the preliminary DMRG run
is sufficiently large that $\ket{\psi(t)}$ can be well approximated 
within that Hilbert space for all times, such that
\begin{equation}
    \epsilon(t) = 1 - | \langle \psi(t) | \psi_{{\rm exact}}(t) 
    \rangle |
\end{equation}
remains small as $t$ grows. This, in general, will only 
be true for relatively short times. A variety of modifications that 
should extend the reach of the static Hilbert space in time can be 
imagined. They typically rest on the DMRG practice of ``targeting'' 
several states: to construct the reduced density matrix used to 
determine the relevant Hilbert space states, one may carry out a 
partial trace over a mixture of a small number of 
states such that the truncated 
Hilbert space is constructed so that all of those states are 
optimally approximated in the DMRG sense:
\begin{equation}
    \hat{\rho}_{S}={\rm Tr}_{E} \ket{\psi}\bra{\psi} \rightarrow
    \hat{\rho}_{S}={\rm Tr}_{E} \sum_{i}\alpha_{i}\ket{\psi_{i}}\bra{\psi_{i}}.     
\end{equation}

A simple choice uses the targeting of $\ham^n \ket{\psi_{0}}$,
for $n$ less than 10 or so, approximating the short-time evolution,
which we have found to substantially improve the 
quality of results for non-adiabatic switching of Hamiltonian 
parameters in time: convergence in $M$ is faster and more consistent
with the new DMRG method (see below). 

Similarly, we have found that for adiabatic changes of Hamiltonian 
parameters results improve if one targets the ground states of both 
the initial and final Hamiltonian.
These approaches are conceptually very similar to targeting not only
$\ket{\psi_{0}}$, but also $\hat{O}\ket{\psi_{0}}$ and some $\ham^n 
\hat{O}\ket{\psi_{0}}$,
$n=1,2,3,\ldots$ in Lanczos vector dynamics DMRG\cite{Hall95,Kuhn99}, or
real and imaginary part of $(\ham-\omega-E_{0}+\imag\eta)^{-1} 
\hat{O}\ket{\psi_{0}}$ in correction vector dynamics DMRG\cite{Kuhn99,Jeck02}
to calculate Green's functions
\begin{equation}
    \bra{\psi_{0}} \hat{O}^\dagger 
    \frac{1}{H-\omega-E_{0}+\imag\eta} \hat{O} \ket{\psi_{0}} .
\end{equation}

\begin{figure}
\centering\epsfig{file=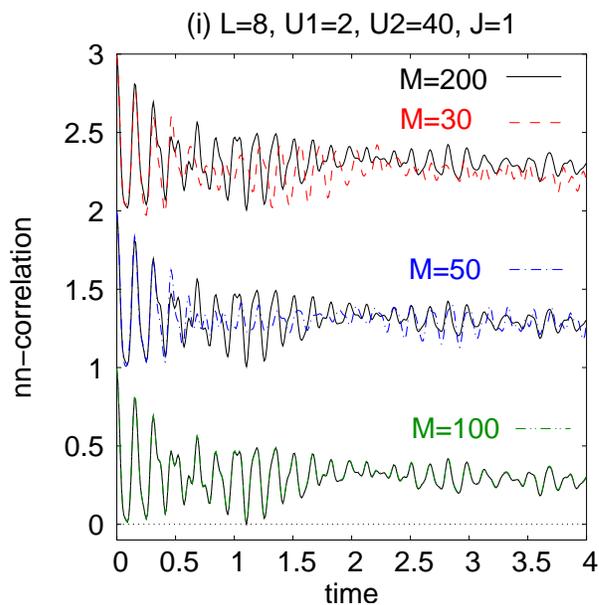,width=0.6\linewidth}
\caption{Time evolution of the real part of the nearest-neighbor correlations in a 
Bose-Hubard model with instantaneous change of interaction strength at $t=0$:
superfluid state targeting only. The different curves for different $M$ are shifted.}
\label{fig:crank0}
\end{figure}

\begin{figure}
\centering\epsfig{file=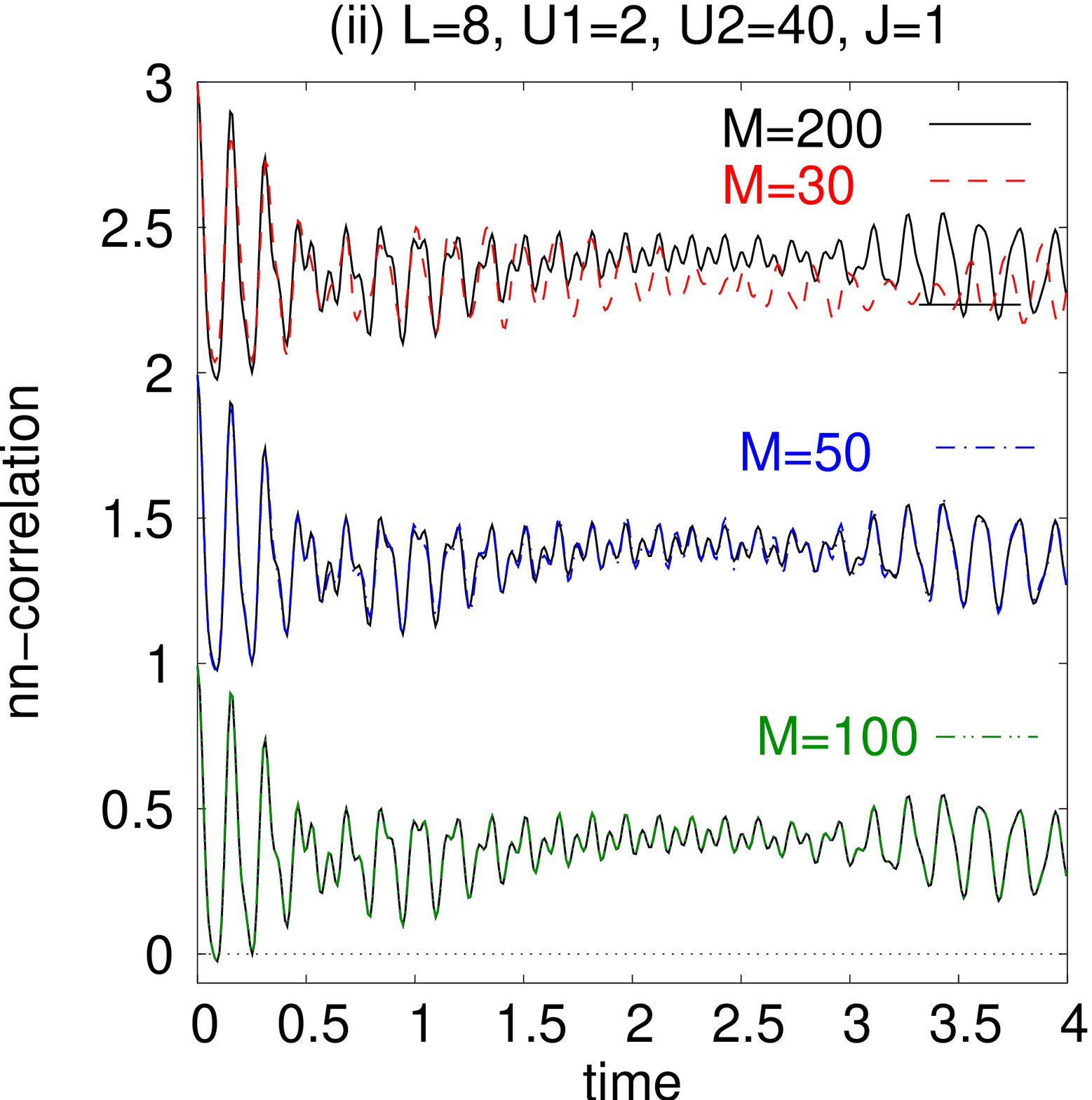,width=0.6\linewidth}
\caption{Time evolution of the real part of the nearest-neighbor correlations in a 
Bose-Hubard model with instantaneous change of interaction strength at $t=0$:
targeting of the initial superfluid ground state, Mott insulating ground state 
and {\em one} time-evolution step. The different curves for different $M$ are shifted.}
\label{fig:crank1}
\end{figure}

\begin{figure}
\centering\epsfig{file=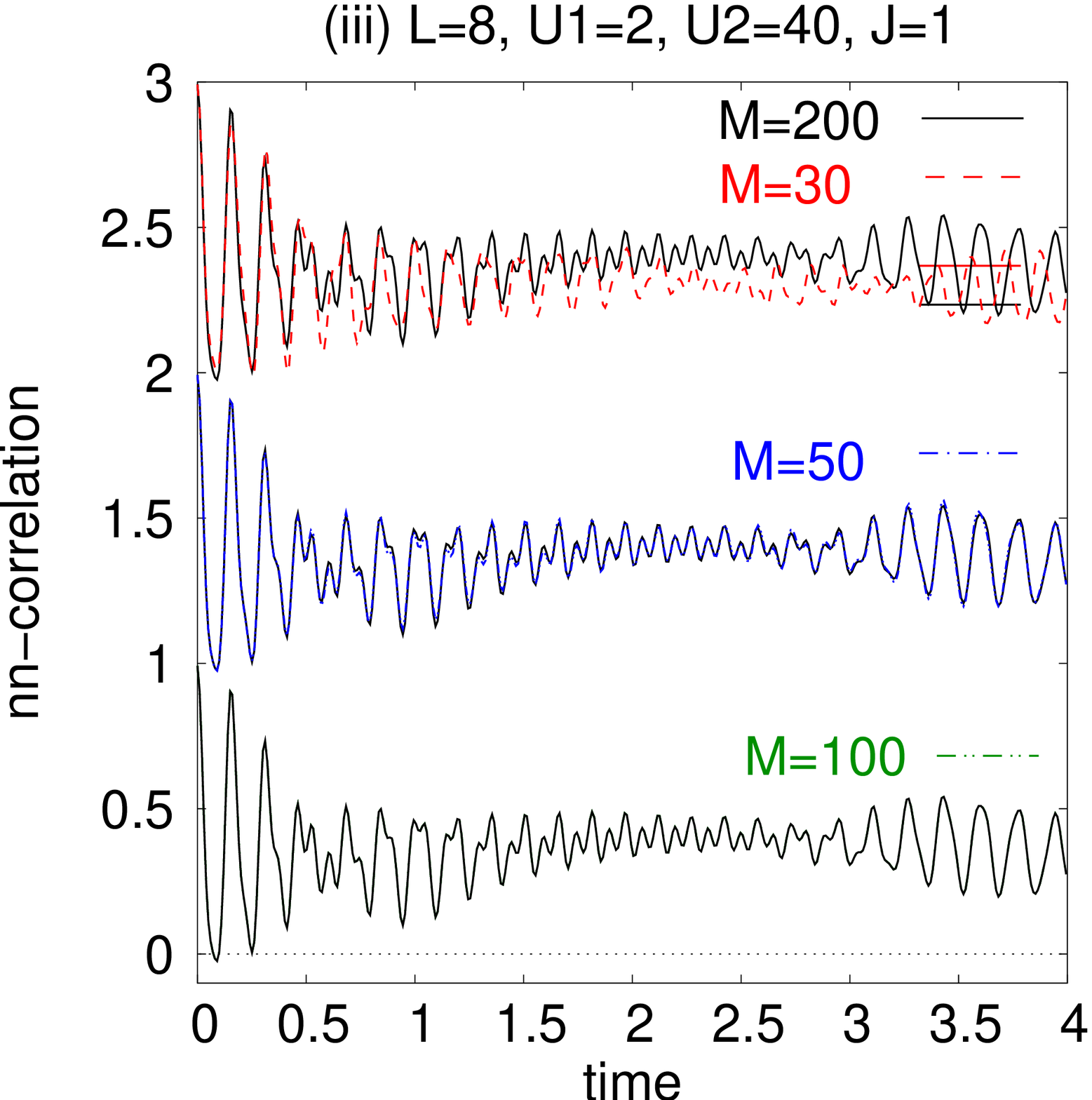,width=0.6\linewidth}
\caption{Time evolution of the real part of the nearest-neighbor correlations in a 
Bose-Hubard model with instantaneous change of interaction strength at $t=0$:
targeting of the initial superfluid ground state, Mott insulating ground state 
and {\em three} time-evolution steps. The different curves for different $M$ are shifted. }
\label{fig:crank3}
\end{figure}

To illustrate the previous approaches, 
we show results for the parameters of the Bose-Hubbard model discussed 
above. Time evolution is calculated in the 
Crank-Nicholson approach using a stepwidth $\delta t = 5 \cdot 
10^{-5}$ in time units of $\hbar/J$
targeting (i) just the superfluid ground state 
$\ket{\psi_{0}}$ for $U_{1}=2$ (Fig. \ref{fig:crank0}), (ii) in addition to (i) also the 
Mott-insulating ground state $\ket{\psi_{0}'}$ for $U_{2}=40$ and
$\ham(t>0)\ket{\psi_{0}}$ (Fig. \ref{fig:crank1}), (iii) in addition to (i) and (ii) also
$\ham(t>0)^2\ket{\psi_{0}}$ and $\ham(t>0)^3\ket{\psi_{0}}$
(Fig. \ref{fig:crank3}). 

We have used up to $M=200$ states to obtain converged 
results (meaning that we could observe no difference between the 
results for $M=100$ and $M=200$) 
for $t\leq 4$, corresponding to roughly 25 oscillations.
The results for the cases (ii) and (iii) are almost converged for $M=50$,
 whereas (i) shows still crude deviations.

\begin{figure}
\centering\epsfig{file=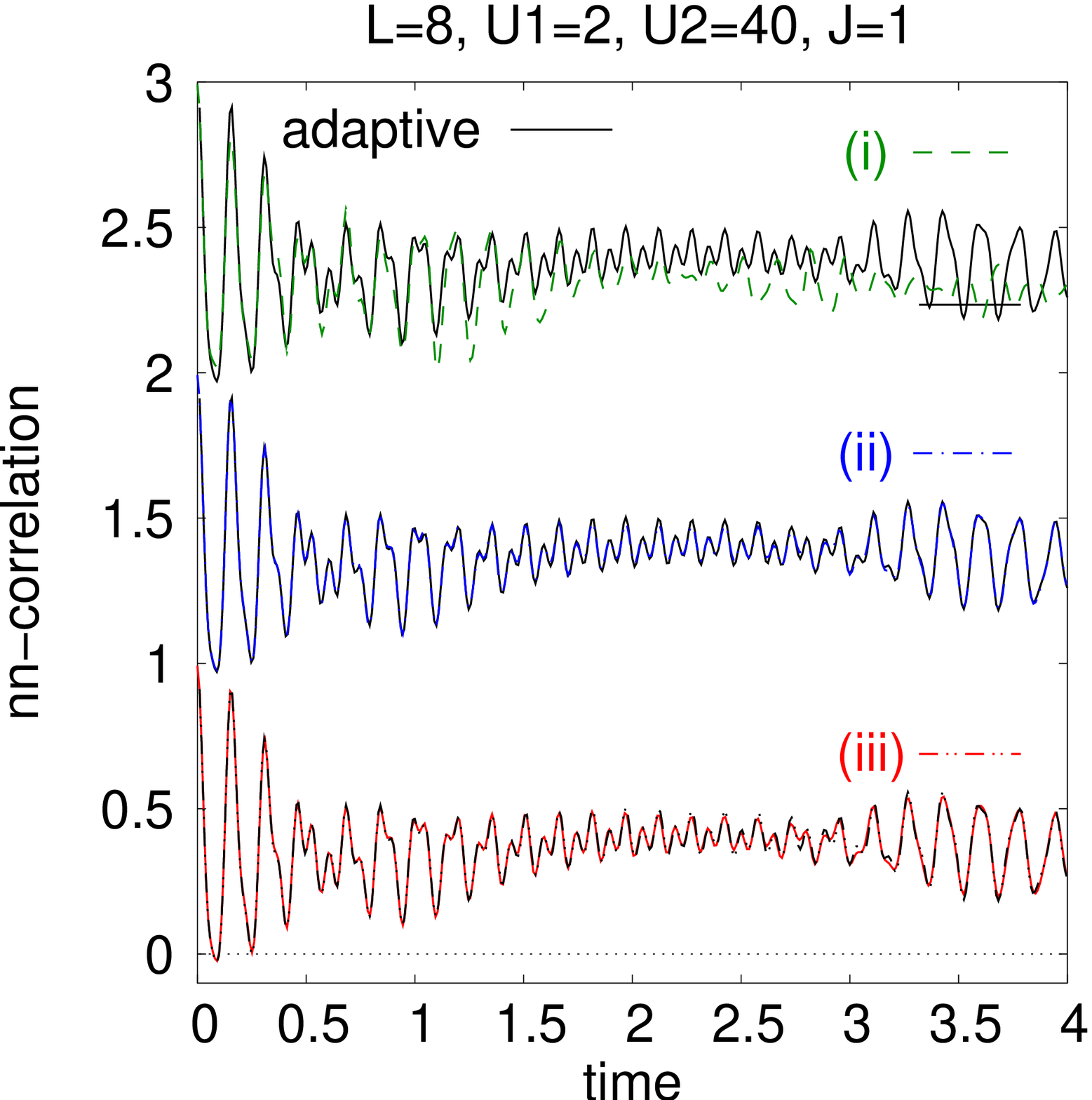,width=0.6\linewidth}
\caption{Comparison of the three $M=200$ Crank-Nicholson calculations to 
adaptive time-dependent DMRG at $M=50$: we target (i) just the superfluid ground state 
$\ket{\psi_{0}}$ for $U_{1}=2$ (Fig. \ref{fig:crank0}), (ii) in addition to (i) also the 
Mott-insulating ground state $\ket{\psi_{0}'}$ for $U_{2}=40$ and
$\ham(t>0)\ket{\psi_{0}}$ (Fig. \ref{fig:crank1}), (iii) in addition to (i) and (ii) also
$\ham(t>0)^2\ket{\psi_{0}}$ and $\ham(t>0)^3\ket{\psi_{0}}$. The different
curves are shifted.}
\label{fig:crankcompn}
\end{figure}

A remarkable observation can be made if one compares the three $M=200$
curves (Fig. \ref{fig:crankcompn}), which by standard DMRG procedure (and for lack of a better 
criterion) would be considered the final, converged outcome, both 
amongst each 
other or to the result of the new adaptive time-dependent DMRG 
algorithm which we are going to discuss below: 
result (i) is clearly {\em not} quantitatively correct 
beyond very short times, whereas result (ii) agrees very 
well with the new algorithm, and result (iii) agrees almost (beside some small
deviations at $t\approx 3$) with result (ii) and the new algorithm.
Therefore we see that for case (i) the criterion of 
convergence in $M$ does not give a good control to
determine if the obtained results are correct. This raises as well doubts about the
reliability of this criterion for cases (ii) and (iii).

A more elaborate, but also much more time-consuming improvement still 
within the framework of a static Hilbert space 
was proposed by 
Luo, Xiang and Wang \cite{Luo03,Caza03}. Additional to the ground state they
target a finite number of quantum states at various discrete times using a 
bootstrap procedure starting from the time evolution of 
smaller systems that are iteratively grown to the desired final size. 

The observation that even relatively robust numerical quantities such as 
nearest-neighbor correlations can be qualitatively and quantitatively 
improved by the additional targeting of states 
which merely share some fundamental characteristics with the true 
quantum state (as we will never reach the Mott-insulating ground 
state) or characterize only the very short-term time evolution
indicates that it would be highly desirable to have a 
modified DMRG algorithm which, for each time $t$, selects Hilbert 
spaces of dimension $M$ such that $\ket{\psi(t)}$ is represented 
optimally in the DMRG sense, 
thus attaining at {\em all} times the typical DMRG precision for 
$M$ retained states. The presentation of such an algorithm is the 
purpose of the following sections.

\section{Matrix product states}
\label{sec:matrixstates}
As both the TEBD simulation algorithm and DMRG can be neatly expressed in the 
language of matrix product states, let us briefly review the properties
of these states also known as finitely-correlated states\cite{Fann92,Klum93}.

We begin by considering a one-dimensional system of size $L$, 
divided up into sites which each have a
local Hilbert space, $\mathcal{H}_i$. For simplicity we  
take the same dimension $N_{{\rm site}}$ at all sites.
In such a system a product state may be expressed as
\begin{equation}
\ket{\fat{\sigma}}=\ket{\sigma_1} \otimes \ket{\sigma_2} \otimes \ldots \otimes
\ket{\sigma_L}, \label{productstate}
\end{equation}
where $\ket{\sigma_i}$ denotes the local state on site $i$. We can express a
general state of the whole system as
\begin{eqnarray}
\ket{\psi}&=&\sum_{\sigma_1,\ldots,\sigma_L}
\psi_{\sigma_1,\ldots,\sigma_L}\ket{\sigma_1}
\otimes \ket{\sigma_2} \otimes \ldots \otimes\ket{\sigma_L} \nonumber\\
&\equiv&\sum_{\fat{\sigma}} \psi_{\fat{\sigma}}\ket{\fat{\sigma}}.
\end{eqnarray}
This general state exists
in the Hilbert space $\mathcal{H}=\prod_{i=1}^L \mathcal{H}_i$, 
with dimension 
$(N_{{\rm site}})^L$.

A matrix product state is now formed by only using a specific set 
of expansion coefficients $\psi_{\fat{\sigma}}$. 
Let us construct this set in the following. To do this we define operators 
$\hat{A}_i[\sigma_i]$ which correspond to a local basis state
$\ket{\sigma_i}$ at site $i$ of the original system, 
but which act on auxiliary spaces of dimension $M$, i.e.,
\begin{equation}
\hat{A}_{i}[\sigma_i]=\sum_{\alpha,\beta}A^i_{\alpha\beta}[\sigma_i]
\ket{\alpha}\bra{\beta},\label{mpsdef}
\label{mpsadef}
\end{equation}
where $\ket{\alpha}$ and $\ket{\beta}$ are orthonormal basis states in auxiliary
spaces. For visualization, we imagine the auxiliary state spaces to be 
located on the bonds next to site $i$. If we label the bond linking 
sites $i$ and $i+1$ by $i$, then we say that the states $\ket{\beta}$ 
live on bond $i$ and the states $\ket{\alpha}$ on bond $i-1$. The 
operators $\hat{A}_i[\sigma_i]$ hence act as transfer operators past 
site $i$ depending on the local state on site $i$. On the first and 
last site, which will need special attention later, this picture 
involves bonds $0$ and $L$ to the left of site 1 and to the right 
of site $L$ respectively. 
While these bonds have no physical meaning for open 
boundary conditions, they are identical and link sites 1 and $L$ as one 
physical bond for 
periodic boundary conditions. There is no a priori 
significance to be attached to the states in the auxiliary state spaces.

In general, operators corresponding to different sites can be different. If
this is the case the
resulting matrix product state to be introduced is referred to as a
position dependent matrix product state. 
We also impose the condition
\begin{equation}
\sum_{\sigma_i} \hat{A}_i[\sigma_i] \hat{A}_i^\dag[\sigma_i]=
\mathcal{I},\label{mpsident}
\end{equation}
which we will see to be related to orthonormality properties of bases 
later.
An unnormalized matrix product state in a form that will be found 
useful for Hamiltonians with open boundary conditions is now 
defined as
\begin{equation}
\ket{\tilde{\psi}}=\sum_{\fat{\sigma}}\left(\bra{\phi_L} \prod_{i=1}^L 
\hat{A}_{i}[\sigma_i]
\ket{\phi_R} \right) \ket{\fat{\sigma}},\label{mpscoef}
\end{equation}
where $\ket{\phi_L}$ and $\ket{\phi_R}$ are the left and right boundary states
in the auxiliary spaces on bonds 0 and $L$. They act on the product of the 
operators $\hat{A}_i$ to produce scalar coefficients
\begin{equation}
    \psi_{\fat{\sigma}} = \bra{\phi_L} \prod_{i=1}^L \hat{A}_{i}[\sigma_i]
\ket{\phi_R}
\end{equation}    
for the expansion of $\ket{\tilde{\psi}}$.

Several remarks are in order.
It should be emphasized that the set of states obeying Eq.\ (\ref{mpscoef}) 
is an (arbitrarily constructed) submanifold of the full 
boundary-condition independent Hilbert space of the quantum many-body 
problem on $L$ sites that is hoped to yield good approximations to 
the true quantum states for Hamiltonians with 
open boundary conditions.
If the dimension $M$ of the auxiliary spaces 
is made sufficiently large then any general
state of the system can, in principle, be represented exactly in this form (provided that
$\ket{\phi_L}$ and $\ket{\phi_R}$ are chosen appropriately), simply 
because the $O(N_{{\rm site}}LM^2)$ degrees of freedom to choose the 
expansion coefficients will exceed $N_{{\rm site}}^L$. This is, of 
course, purely academic. The practical relevance of the matrix product states
even for computationally manageable values of $M$ is shown by the success of
DMRG, which is known\cite{Ostl95,Duke98} to produce 
matrix product states of auxiliary state space dimension $M$, in 
determining energies and correlators at very high precision for 
moderate values of $M$. In fact, some very important 
quantum states in one dimension, such as the valence-bond-solid (VBS) ground 
state of the Affleck-Kennedy-Lieb-Tasaki (AKLT) model 
\cite{Affl87,Affl88,Fann89}, can be described exactly by matrix 
product states using very small $M$ ($M=2$ for the AKLT model).  

Let us now formulate a Schmidt decomposition for matrix product states which
 can be done very easily.
An unnormalized state $\ket{\tilde{\psi}}$ 
of the matrix-product form of Eq.\ 
(\ref{mpscoef}) with auxiliary space dimension $M$ can be written as 
\begin{equation}
    \ket{\tilde{\psi}}= \sum_{\alpha=1}^{M} \ket{\tilde{w}^{S}_{\alpha}} 
    \ket{\tilde{w}^{E}_{\alpha}} ,
    \label{eq:decomp18}
\end{equation}
where we have arbitrarily cut the chain into S on the left 
and E on the right with
\begin{equation}
\ket{\tilde{w}^{S}_{\alpha}} = \sum_{\{\fat{\sigma}^S\}} \left[ 
\bra{\phi_{L}} 
 \prod_{i\in S} \hat{A}_{i} [\sigma_{i}] \ket{\alpha} \right] 
\ket{\fat{\sigma}^S} ,
    \label{eq:mpshalf}
\end{equation}
and similarly $\ket{\tilde{w}^{E}_{\alpha}}$, where $\{ \ket{\alpha} 
\}$ are the states spanning the auxiliary state space on the cut bond. 
Normalizing the states 
$\ket{\tilde{\psi}}$,
$\ket{\tilde{w}^{S}_{\alpha}}$ and $\ket{\tilde{w}^{E}_{\alpha}}$ we obtain
the representation   
\begin{equation}
 \ket{\psi} = \sum_{\alpha=1}^{M} 
    \lambda_{\alpha}\ket{w^{S}_{\alpha}} 
    \ket{w^{E}_{\alpha}} 
\end{equation}
where in $\lambda_{\alpha}$ the factors resulting from the normalization are
absorbed.
 The relationship 
to reduced density matrices is as detailed in Sec.\ \ref{sec:intro}.

\section{TEBD Simulation Algorithm}
\label{sec:vidal}
Let us now express the TEBD simulation algorithm in the language of the previous 
section. In the original exposition of the algorithm \cite{Vida03a}, 
one starts from a representation of a quantum state where 
the coefficients for the
states are decomposed as a product of tensors,
\begin{equation}
\psi_{\sigma_1,\ldots,\sigma_L}=\sum_{\alpha_1,\ldots,\alpha_{L-1}} \Gamma^{[1]
\sigma_1}_{\alpha_1} \lambda^{[1]}_{\alpha_1} \Gamma^{[2] \sigma_2}_{\alpha_1
\alpha_2}\lambda^{[2]}_{\alpha_2} \Gamma^{[3] \sigma_3}_{\alpha_2 \alpha_3}\cdots
\Gamma^{[L] \sigma_L}_{\alpha_{L-1}}.\label{vidalcoef}
\end{equation}

It is of no immediate concern to us how the
$\Gamma$ and $\lambda$ tensors are constructed explicitly for a given 
physical situation. Let us assume that they have been determined such 
that they approximate the true wave function close to the optimum 
obtainable within the class of wave functions having such
coefficients; this is indeed possible as will be discussed below. 
There are, in fact, two ways of doing it, within the framework of 
DMRG (see below), or by a continuous imaginary time 
evolution from some simple product state, as discussed in Ref.\ 
\cite{Vida03b}. 

Let us once again attempt a visualization; the 
(diagonal) tensors $\lambda^{[i]}$, $i=1,\ldots,L-1$ are associated 
with the bonds $i$, whereas $\Gamma^{[i]}$, $i=2,\ldots,L-1$ links 
(transfers) from bond $i$ to bond $i-1$ across site $i$. Note that at 
the boundaries ($i=1,L$) the structure of the $\Gamma$ is different, 
a point of importance in the following. The sums run 
over $M$ states $\ket{\alpha_{i}}$ living in auxiliary state spaces on bond $i$.
A priori, these states have no physical meaning here.

\begin{figure}
\centering\epsfig{file=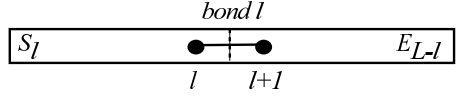,scale=0.9}
\caption{Bipartitioning by cutting bond $l$ between sites $l$ and $l+1$.}
\label{fig:cut}
\end{figure}

The $\Gamma$ and $\lambda$ tensors are constructed such that for an
arbitrary cut of the system into a part $S_{l}$ of length $l$
and a part $E_{L-l}$ of length $L-l$ at bond $l$,
the Schmidt decomposition for this bipartite splitting reads
\begin{equation}
\ket{\psi}=\sum_{\alpha_l}\lambda^{[l]}_{\alpha_l}\ket{w_{\alpha_l}^{S_{l}}}
\ket{w_{\alpha_l}^{E_{L-l}}},
\end{equation}
with
\begin{equation}
\ket{w_{\alpha_l}^{S_{l}}}= \sum_{\alpha_1,\ldots,\alpha_{l-1}}
\sum_{\sigma_1,\ldots,\sigma_{l}} \Gamma^{[1]
\sigma_1}_{\alpha_1} \lambda^{[1]}_{\alpha_1} \cdots \Gamma^{[l]
\sigma_l}_{\alpha_{l-1}\alpha_l} 
\ket{\sigma_1}\otimes\cdots\otimes 
\ket{\sigma_l},
\label{eq:growsys}
\end{equation}
and
\begin{eqnarray}
\ket{w_{\alpha_l}^{E_{L-l}}}&=& \sum_{\alpha_l,\ldots,\alpha_{L-1}}
\sum_{\sigma_{l+1},\ldots,\sigma_{L}} \Gamma^{[l+1]
\sigma_{l+1}}_{\alpha_l \alpha_{l+1}}\lambda^{[l+1]}_{\alpha_{l+1}}  \cdots \Gamma^{[L]
\sigma_L}_{\alpha_{L-1}} \times \nonumber \\
& & \ket{\sigma_{l+1}}\otimes\cdots\otimes
\ket{\sigma_L},\label{vidalphibdef}
\end{eqnarray}
where $\ket{\psi}$ is normalized and the sets of 
$\{\ket{w_{\alpha_l}^{S_{l}}} \}$ and $\{ \ket{w_{\alpha_l}^{E_{L-l}}}\}$
are orthonormal. This implies, for example, that
\begin{equation}
    \sum_{\alpha_{l}} (\lambda_{\alpha_{l}}^{[l]})^2 = 1.
\end{equation}    

We can see that (leaving aside normalization considerations for the 
moment) this representation may be expressed as a matrix product state if we
choose for
$\hat{A}_i[\sigma_i]=\sum_{\alpha,\beta}A^i_{\alpha\beta}[\sigma_i]\ket{\alpha}\bra{\beta}$
\begin{equation}
A^i_{\alpha\beta}[\sigma_i]=\Gamma^{[i]\sigma_i}_{\alpha\beta}\lambda^{[i]}_\beta,
\end{equation}
except for $i=1$, where we choose
\begin{equation}
A^1_{\alpha\beta}[\sigma_1]=f_\alpha\Gamma^{[1]\sigma_1}_{\beta}\lambda^{[1]}_\beta,
\end{equation}
and for $i=L$, where we choose
\begin{equation}
A^L_{\alpha\beta}[\sigma_L]=\Gamma^{[L]\sigma_L}_{\alpha} g_\beta.
\end{equation}
The vectors $f_\alpha$ and $g_\beta$ are normalised vectors which must be chosen
in conjunction with the boundary states $\ket{\phi_L}$ and $\ket{\phi_R}$ so as to
produce the expansion (\ref{vidalcoef}) from this choice of the $\hat{A}_i$.
Specifically, we require
\begin{eqnarray}
\ket{\phi_L}&=&\sum_\alpha f_\alpha \ket{\alpha}\\
\ket{\phi_R}&=&\sum_\beta g^*_\beta \ket{\beta},
\end{eqnarray}
where $\ket{\alpha}$ and $\ket{\beta}$ are the states forming the same orthonormal basis
in the auxiliary spaces on bonds 0 and $L$ 
used to express $A^i_{\alpha\beta}$. In typical implementations of
the algorithm it is common to take $f_\alpha=g_\alpha=\delta_{\alpha,1}$.
Throughout the rest of the article we take this as the definition for
$g_\alpha$ and $f_\alpha$, as this allows us to treat the operators on the 
boundary identically to the other operators for the purposes of the simulation protocol.
For the same reason we define a vector $\lambda^{[0]}_\alpha=\delta_{\alpha,1}$.

In the above expression we have grouped $\Gamma$ and $\lambda$ such 
that the $\lambda$ reside on the {\em right} of the two bonds linked 
by $\Gamma$. There is another valid choice for the $\hat{A}_i$, 
which will produce identical states in the original system, and essentially 
the same procedure for the algorithm. If we set
\begin{equation}
\tilde{A}^i_{\alpha\beta}[\sigma_i]=\lambda^{[i-1]}_\alpha\Gamma^{[i]\sigma_i}_{\alpha\beta},
\end{equation}
except for $i=1$, where we choose
\begin{equation}
\tilde{A}^1_{\alpha\beta}[\sigma_1]=f_\alpha\Gamma^{[1]\sigma_1}_{\beta},
\end{equation}
and for $i=L$, where we choose
\begin{equation}
\tilde{A}^L_{\alpha\beta}[\sigma_L]=\lambda^{[L-1]}_\alpha \Gamma^{[L]\sigma_L}_{\alpha}
g_\beta,
\end{equation}
then the same choice of boundary states produces the correct coefficients. 
Here we have grouped $\Gamma$ and $\lambda$ such 
that the $\lambda$ reside on the {\em left} of the two bonds linked 
by $\Gamma$.  It is also important to note that 
any valid choice of $f_\alpha$ and $g_\beta$ that produces the expansion
(\ref{vidalcoef}) specifically {\em excludes} the use of periodic boundary conditions.
While generalizations are feasible, they lead to a much more 
complicated formulation of the TEBD simulation algorithm and will not 
be pursued here.

To conclude the identification of states, let us consider 
normalization issues. The condition (\ref{mpsident}) is indeed fulfilled 
for our choice of $A_i[\sigma_i]$, because we
have from (\ref{vidalphibdef}) for a splitting at $l$ that
\begin{eqnarray}
\ket{w_{\alpha_{l-1}}^{E_{L-(l-1)}}}&=&\sum_{\alpha_{l}\sigma_{l}}\Gamma^{[l]
\sigma_l}_{\alpha_{l-1} \alpha_{l}} \lambda^{[l]}_{\alpha_{l}} \ket{\sigma_{l}} \otimes
\ket{w_{\alpha_{l}}^{E_{L-l}}}\nonumber\\
&=&\sum_{\alpha_{l}\sigma_{l}} A^l_{\alpha_{l-1}\alpha_l}[\sigma_l]\ket{\sigma_{l}} \otimes
\ket{w_{\alpha_{l}}^{E_{L-l}}},
\label{eq:recursion}
\end{eqnarray}
so that from the orthonormality of the sets of states
$\{\ket{w_{\alpha}^{E_{L-(l-1)}}}\}_{\alpha=1}^M$, 
$\{\ket{\sigma_{l}}\}_{\sigma_l=1}^{N_{{\rm site}}}$
and $\{\ket{w_{\gamma}^{E_{L-l}}}\}_{\gamma=1}^M$,
\begin{eqnarray}
\sum_{\sigma_l} \hat{A}_l[\sigma_l] \hat{A}^\dag_l[\sigma_l]
&=&\sum_{\alpha\beta\gamma}\sum_{\sigma_l} A^l_{\alpha\gamma}[\sigma_l] (A^l_{\beta
\gamma}[\sigma_l])^* \ket{\alpha}\bra{\beta}
\nonumber\\
&=&\sum_{\alpha\beta}\bra{w_\beta^{E_{L-(l-1)}}}w_\alpha^{E_{L-(l-1)}}\rangle 
\ket{\alpha}\bra{\beta}
\nonumber\\
&=&\sum_{\alpha\beta} \delta_{\alpha \beta} \ket{\alpha}\bra{\beta} = \mathcal{I}.
\end{eqnarray}
Let us now consider the time evolution for a typical (possibly time-dependent)
Hamiltonian in 
strongly correlated systems that contains only short-ranged 
interactions, for simplicity only nearest-neighbor interactions here:
\begin{equation}
\hat{H}=\sum_{i \: \rm odd}\hat{F}_{i,i+1} + \sum_{j \rm \: even}\hat{G}_{j,j+1},
\end{equation}
$F_{i,i+1}$ and $G_{j,j+1}$ are the local Hamiltonians on the odd bonds 
linking $i$ and $i+1$, and the even bonds linking $j$ and $j+1$. While all
$F$ and $G$ terms commute among each other, $F$ and $G$ terms 
do in general not commute if they share one site.
Then the time evolution operator may be approximately represented by a (first order)
Trotter expansion as
\begin{equation}
\rmeh^{-\rmih \hat{H} \delta t}=\prod_{i \rm \: odd} \rmeh^{-\rmih \hat{F}_{i,i+1} \delta
t}\prod_{j \rm \: even} \rmeh^{-\rmih \hat{G}_{j,j+1} \delta t} + \mathcal{O}(\delta t^2),
\end{equation}
and the time evolution of the state can be computed by repeated application of the
two-site time evolution operators $\exp({-\rmih \hat{G}_{j,j+1} \delta t})$ and
$\exp({-\rmih \hat{F}_{i,i+1} \delta t})$. This is a well-known 
procedure in particular in Quantum Monte Carlo\cite{Suzu76} where it serves to 
carry out imaginary time evolutions (checkerboard decomposition). 

The TEBD simulation algorithm now runs as follows\cite{Vida03a,Vida03b}:
\begin{enumerate}
    \item Perform the following two steps for all even bonds (order does not
      matter):
\begin{itemize}
     \item[(i)] Apply $\exp({-\rmih \hat{G}_{l,l+1} \delta 
    t})$ to $\ket{\psi(t)}$. For each local time update, a new wave function is 
    obtained. The number of degrees of 
    freedom on the ``active'' bond thereby increases, as will be detailed below.  
    \item[(ii)] Carry out a Schmidt 
    decomposition cutting this bond and retain as in DMRG only 
    those $M$ degrees of freedom with the highest weight in the 
    decomposition. 
\end{itemize}
    \item Repeat this two-step procedure for all {\em odd} bonds, 
    applying $\exp({-\rmih \hat{F}_{l,l+1} \delta 
    t})$.
    \item This completes one Trotter time step. One may now evaluate 
    expectation values at selected time steps, and continues the 
    algorithm from step 1.
\end{enumerate}

Let us now consider the computational details.
\newline
(i) Consider a local time evolution operator acting on bond $l$, i.e.\ 
sites $l$ and $l+1$, for a state $\ket{\psi}$. The Schmidt 
decomposition of $\ket{\psi}$ after partitioning by cutting bond $l$ 
reads
\begin{equation}
    \ket{\psi}=\sum_{\alpha_{l}=1}^M \lambda^{[l]}_{\alpha_{l}}
    \ket{w_{\alpha_{l}}^{S_{l}}} \ket{w_{\alpha_{l}}^{E_{L-l}}} .
\label{oldstate}
\end{equation}
Using Eqs.\ (\ref{eq:growsys}), (\ref{vidalphibdef}) and 
(\ref{eq:recursion}), we find
\begin{eqnarray}
    \ket{\psi}&=& \sum_{\alpha_{l-1}\alpha_{l}\alpha_{l+1}} 
    \sum_{\sigma_{l}\sigma_{l+1}} 
    \lambda^{[l-1]}_{\alpha_{l-1}} 
    A^{l}_{\alpha_{l-1}\alpha_{l}}[\sigma_{l}] 
    A^{l+1}_{\alpha_{l}\alpha_{l+1}}[\sigma_{l+1}] \times \nonumber \\
    & & \ket{w_{\alpha_{l-1}}^{S_{l-1}}} \ket{\sigma_{l}} \ket{\sigma_{l+1}}
    \ket{w_{\alpha_{l+1}}^{E_{L-(l+1)}}} .    
\end{eqnarray}    
We note, that if we identify $\ket{w_{\alpha_{l-1}}^{S_{l-1}}}$ and
$\ket{w_{\alpha_{l+1}}^{E_{L-(l+1)}}}$ with DMRG system and environment 
block states $\ket{w^S_{m_{l-1}}}$ and $\ket{w^E_{m_{l+1}}}$, 
we have a typical DMRG state for two blocks and two sites
\begin{equation}
    \ket{\psi}=
    \sum_{m_{l-1}} 
    \sum_{\sigma_{l}} 
    \sum_{\sigma_{l+1}} 
    \sum_{m_{l+1}} \psi_{m_{l-1}\sigma_{l}\sigma_{l+1} m_{l+1}} 
    \ket{w^S_{m_{l-1}}}\ket{\sigma_{l}}\ket{\sigma_{l+1}}\ket{w^E_{m_{l+1}}} 
\end{equation}
with
\begin{equation}
    \psi_{m_{l-1} \sigma_{l}\sigma_{l+1} m_{l+1}} =
    \sum_{\alpha_{l}} \lambda^{[l-1]}_{m_{l-1}} 
    A^{l}_{m_{l-1}\alpha_{l}}[\sigma_{l}] 
    A^{l+1}_{\alpha_{l}m_{l+1}}[\sigma_{l+1}] .
\end{equation}

The local time evolution operator on site $l, l+1$ can be expanded as  
\begin{equation}
    \hat{U}_{l,l+1} = \sum_{\sigma_{l}\sigma_{l+1}} 
    \sum_{\sigma_{l}'\sigma_{l+1}'} 
    U^{\sigma_{l}'\sigma_{l+1}'}_{\sigma_{l}\sigma_{l+1}}
    \ket{\sigma_{l}'\sigma_{l+1}'}\bra{\sigma_{l}\sigma_{l+1}}
\end{equation}
and generates $\ket{\psi'} = \hat{U}_{l,l+1} \ket{\psi}$, where
\begin{eqnarray*}
    \ket{\psi'} &=& \sum_{\alpha_{l-1}\alpha_{l}\alpha_{l+1}} 
    \sum_{\sigma_{l}\sigma_{l+1}} 
    \sum_{\sigma_{l}'\sigma_{l+1}'} \\
    & & \lambda^{[l-1]}_{\alpha_{l-1}} 
    A^{l}_{\alpha_{l-1}\alpha_{l}}[\sigma_{l}'] 
    A^{l+1}_{\alpha_{l}\alpha_{l+1}}[\sigma_{l+1}'] U^{\sigma_{l}\sigma_{l+1}}_{\sigma_{l}'\sigma_{l+1}'} 
    \ket{w_{\alpha_{l-1}}^{S_{l-1}}} \ket{\sigma_{l}} \ket{\sigma_{l+1}}
    \ket{w_{\alpha_{l+1}}^{E_{L-(l+1)}}} . 
\end{eqnarray*}

This can also be written as 
\begin{equation}
    \ket{\psi'}= \sum_{\alpha_{l-1}\alpha_{l+1}} 
    \sum_{\sigma_{l}\sigma_{l+1}} 
    \Theta^{\sigma_{l}\sigma_{l+1}}_{\alpha_{l-1}\alpha_{l+1}} 
    \ket{w_{\alpha_{l-1}}^{S_{l-1}}} \ket{\sigma_{l}} \ket{\sigma_{l+1}}
    \ket{w_{\alpha_{l+1}}^{E_{L-(l+1)}}} ,
\end{equation}    
where
\begin{equation}
    \Theta^{\sigma_{l}\sigma_{l+1}}_{\alpha_{l-1}\alpha_{l+1}}
=   \lambda^{[l-1]}_{\alpha_{l-1}}  \sum_{\alpha_{l}\sigma_{l}'\sigma_{l+1}'}
    A^{l}_{\alpha_{l-1}\alpha_{l}}[\sigma_{l}'] 
        A^{l+1}_{\alpha_{l}\alpha_{l+1}}[\sigma_{l+1}'] 
        U^{\sigma_{l}\sigma_{l+1}}_{\sigma_{l}'\sigma_{l+1}'} . 
\end{equation}
(ii) Now a {\em new} Schmidt decomposition identical to that in DMRG 
can be carried out for 
$\ket{\psi'}$: cutting once again bond $l$, there are now 
$MN_{{\rm site}}$ states in each part of the system, leading to 
\begin{equation}
    \ket{\psi'}=\sum_{\alpha_{l}=1}^{MN_{{\rm site}}} 
    \tilde{\lambda}^{[l]}_{\alpha_{l}}
    \ket{\tilde{w}_{\alpha_{l}}^{S_{l}}} 
    \ket{\tilde{w}_{\alpha_{l}}^{E_{L-l}}} .
\end{equation}
In general the states and 
coefficients of the decomposition will have changed compared to the 
decomposition (\ref{oldstate}) previous to the time evolution, and hence they are {\em 
adaptive}. We indicate this by introducing a tilde for these states 
and coefficients.
As in DMRG, if there are more than $M$ non-zero eigenvalues, we now choose the 
$M$ eigenvectors corresponding to the largest $\tilde{\lambda}^{[l]}_{\alpha_{l}}$ 
to use in these expressions. The error in the final state produced as a result 
is proportional to the sum of the magnitudes of the discarded 
eigenvalues. After normalization, to allow for the discarded 
weight, the state reads
\begin{equation}
    \ket{\psi'}=\sum_{\alpha_{l}=1}^{M} 
    \lambda^{[l]}_{\alpha_{l}}
    \ket{w_{\alpha_{l}}^{S_{l}}} 
    \ket{w_{\alpha_{l}}^{E_{L-l}}}.
\end{equation}
Note again that the states and coefficients in this superposition are in general different from 
those in Eq.\ (\ref{oldstate}); we have now dropped the tildes again, as 
this superposition will be the starting point for the next time 
evolution (state adaption) step.
As is done in DMRG, to obtain the Schmidt decomposition reduced density 
matrices are formed, e.g.\
\begin{eqnarray}
\dm_{E}&=&{\rm Tr}_{S} \ket{\psi^\prime}\bra{\psi^\prime}\nonumber\\
&=&\sum_{\sigma_{l+1} \sigma_{l+1}^\prime \alpha_{l+1} \alpha_{l+1}^\prime } 
\ket{\sigma_{l+1}} \ket{w_{\alpha_{l+1}}} \bra{w_{\alpha^\prime_{l+1}}} 
\bra{\sigma_{l+1}^\prime}\nonumber \\
& & \times \left(\sum_{\alpha_{l-1}\sigma_l} \Theta^{\sigma_l 
\sigma_{l+1}}_{\alpha_{l-1}\alpha_{l+1}}
(\Theta^{\sigma_l \sigma_{l+1}^\prime}_{\alpha_{l-1}\alpha_{l+1}^\prime})^*
\right) .
\end{eqnarray}

If we now diagonalise $\dm_{E}$, we can read off the new values of 
$A^{l+1}_{\alpha_{l} \alpha_{l+1}}[\sigma_{l+1}]$ 
because the eigenvectors $\ket{w_{\alpha_{l}}^{E_{L-l}}}$ obey
\begin{equation}
    \ket{w_{\alpha_{l}}^{E_{L-l}}} =    
    \sum_{\sigma_{l+1}\alpha_{l+1}}A^{l+1}_{\alpha_{l}\alpha_{l+1}}[\sigma_{l+1}]
    \ket{\sigma_{l+1}}\ket{w_{\alpha_{l+1}}^{E_{L-(l+1)}}}.
\end{equation}
We also obtain the eigenvalues, $(\lambda_{\alpha_{l}}^{[l]})^2$.
Due to the asymmetric grouping of $\Gamma$ and $\lambda$ into $A$ 
discussed above, a short calculation shows that the new values for 
$A^{l}_{\alpha_{l-1} \alpha_{l}}[\sigma_l]$ can be read off from the 
slightly more complicated expression
\begin{equation}
\lambda^{[l]}_{\alpha_{l}} \ket{w^{S_{l}}_{\alpha_{l}}} =
\sum_{\alpha_{l-1}\sigma_{l}} \lambda^{[l-1]}_{\alpha_{l-1}} 
A^{l}_{\alpha_{l-1}\alpha_{l}}[\sigma_{l}] \ket{w^{S_{l-1}}_{\alpha_{l-1}}} \ket{\sigma_{l}} .
\end{equation}
The states $\ket{w^{S_{l}}_{\alpha_{l}}}$ are the normalized 
eigenvectors of $\dm_{S}$ formed in analogy to $\dm_{E}$.

The key point about the TEBD simulation algorithm is that a 
DMRG-style truncation to 
keep the most relevant density matrix eigenstates (or the maximum 
amount of entanglement) is carried out {\em at each time step.} This is in contrast
with time-dependent DMRG methods up to now, 
where the basis states were chosen before the time
evolution, and did not ``adapt'' to optimally represent the final state.

\section{DMRG and matrix-product states} 
\begin{figure}
\centering\epsfig{file=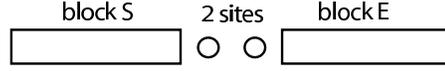,scale=0.9}
\caption{Typical two-block two-site setup of DMRG as used here.}
\label{fig:dmrgsetup}
\end{figure}

Typical normalized DMRG states for the combination of two blocks S and E and two 
single sites (Fig. \ref{fig:dmrgsetup}) have the form
\label{sec:dmrg}
\begin{equation}
    \ket{\psi} =
    \sum_{m_{l-1}} 
    \sum_{\sigma_{l}} 
    \sum_{\sigma_{l+1}} 
    \sum_{m_{l+1}} \psi_{m_{l-1}\sigma_{l}\sigma_{l+1} m_{l+1}} 
    \ket{w^S_{m_{l-1}}}\ket{\sigma_{l}}\ket{\sigma_{l+1}}\ket{w^E_{m_{l+1}}} 
    \label{eq:be-ground}
\end{equation}
which can be Schmidt decomposed as
\begin{equation}
    \ket{\psi}=\sum_{m_{l}} \lambda^{[l]}_{m_{l}} \ket{w^S_{m_{l}}}
    \ket{w^E_{m_{l}}} .
\end{equation}    

It has been known for a long time\cite{Ostl95,Duke98} that a DMRG calculation
retaining $M$ block states produces $M\times M$ matrix-product states 
for $\ket{\psi}$.
Consider the reduced basis transformation to obtain the states of
DMRG block S that terminates on bond $l$ from those of the block terminating 
on bond $l-1$ and those on a single site $l$,
\begin{equation}
    \langle w^S_{m_{l-1}}\sigma_{l} | w^{S}_{m_{l}} \rangle \equiv 
     A^{l}_{m_{l-1}m_{l}}[\sigma_{l}],
\label{eq:indexedtrafomatrix}
 \end{equation}    
such that 
\begin{equation}
    \ket{w^S_{m_{l}}} = \sum_{m_{l-1}\sigma_{l}} 
    A^{l}_{m_{l-1}m_{l}}[\sigma_{l}]\ket{w^S_{m_{l-1}}}\otimes\ket{\sigma_{l}}.
    \label{eq:indexedtrafo}
\end{equation}    
The reduced basis transformation matrices $A_{l}[\sigma_{l}]$ 
automatically obey Eq.\ 
(\ref{mpsident}), which here ensures that $\{ \ket{w^S_{m_{l}}}\}$ is 
an orthonormal set provided $\{ \ket{w^S_{m_{l-1}}}\}$ is one, too.
We may now use Eq.\ (\ref{eq:indexedtrafo}) for a backward recursion 
to express $\ket{w^S_{m_{l-1}}}$ via $\ket{w^S_{m_{l-2}}}$ and so forth. 
There is a complication as the number of 
block states for very short blocks is less than $M$. For simplicity, 
we assume that $M$ is chosen such that we have exactly 
$N_{{\rm site}}^{\tilde{N}}=M$. If we stop the recursion at the shortest block 
of size $\tilde{N}$ that has $M$ states we obtain
\begin{eqnarray}
    \ket{w^S_{m_{l}}} &=& \sum_{m_{\tilde{N}+1}\ldots m_{l-1}}
    \sum_{\sigma_{1}\ldots\sigma_{l}} \\
    & & A^{\tilde{N}+1}_{m_{\tilde{N}}m_{\tilde{N}+1}} 
    [\sigma_{\tilde{N}+1}] 
    \ldots A^l_{m_{l-1}m_{l}} [\sigma_{l}]  
    \ket{\sigma_{1}\ldots\sigma_{l}}, \nonumber
\end{eqnarray}    
where we have boundary-site states on the first $\tilde{N}$ sites indexed by
$m_{\tilde{N}}\equiv\{ \sigma_{1}\ldots\sigma_{\tilde{N}} \}$.

Similarly, for the DMRG block E we have
\begin{equation}
    \langle w^E_{m_{l+1}}\sigma_{l+1} | w^{E}_{m_{l}} \rangle \equiv 
     A^{l+1}_{m_{l}m_{l+1}}[\sigma_{l+1}],
\label{eq:indexedtrafomatrix1}
 \end{equation}    
such that (again having $\tilde{N}$ boundary sites) a recursion gives
\begin{eqnarray}
    \ket{w^E_{m_{l}}} &=& \sum_{m_{l+1}\ldots m_{L-\tilde{N}}}
    \sum_{\sigma_{l+1}\ldots\sigma_{L}} \nonumber \\
    & & A^{l+1}_{m_{l}m_{l+1}} [\sigma_{l+1}]
    \ldots A^{L-\tilde{N}}_{m_{L-\tilde{N}-1}m_{L-\tilde{N}}} 
    [\sigma_{L-\tilde{N}}]       
    \ket{\sigma_{l+1}\ldots\sigma_{L}},
\end{eqnarray}    
with boundary-site states on the last $\tilde{N}$ sites indexed by
$m_{L-\tilde{N}}\equiv\{ \sigma_{L-\tilde{N}+1}\ldots\sigma_L \}$.

A comparison with Eqs.\
(\ref{mpscoef}), (\ref{eq:decomp18}) and (\ref{eq:mpshalf}) 
shows that DMRG generates position-dependent 
$M\times M$ matrix-product states as block states for a reduced 
Hilbert space of $M$ states; the auxiliary state space to a bond 
is given by the Hilbert space of the block at whose end the bond sits. 
This physical meaning attached to the 
auxiliary state spaces and the fact that for the shortest block
the states can be labeled by good quantum numbers (if available) ensures 
through (\ref{eq:indexedtrafomatrix}) and
(\ref{eq:indexedtrafomatrix1})
that they carry good quantum numbers for {\em all} block sizes. The big 
advantage is that using good quantum numbers allows us to exclude a 
large amount of wave function coefficients as being 0, drastically speeding
up all calculations by at least one, and often two orders of magnitude. 
Moreover, as is well known, DMRG can be easily 
adapted to periodic boundary conditions, which is in principle also 
possible for the TEBD algorithm but cumbersome to implement. 
Fermionic degrees of freedom 
also present no specific problem, and in particular, there exists 
no negative sign problem of the kind that is present in Quantum Monte Carlo
methods.

\begin{figure}
\centering\epsfig{file=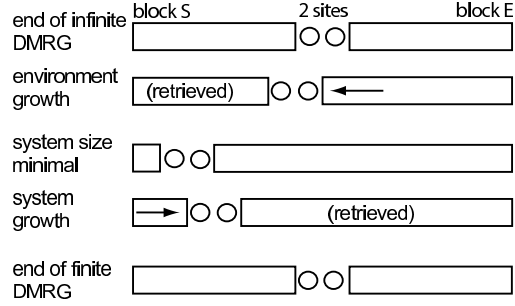,scale=0.72}
\caption{Finite-system DMRG algorithm. Block growth and shrinkage. 
For the adaptive time-dependent DMRG, replace ground state 
optimization by local time evolution.}
\label{fig:finitesize}
\end{figure}

The effect of the finite-system DMRG algorithm\cite{Whit93} is now to shift the two 
free sites through the chain, growing and shrinking the blocks S and E as
illustrated in Fig. \ref{fig:finitesize}.
At each step, the ground state is redetermined and a new Schmidt 
decomposition carried out in which the system is cut between the two free sites, leading 
to a new truncation and new reduced basis transformations (2 matrices 
$A$ adjacent to this bond). It is thus a sequence of {\em local} optimization 
steps of the wave function oriented towards an optimal representation 
of the ground state. Typically, after some ``sweeps'' of the free 
sites from left to right and back, physical quantities evaluated for 
this state converge. While comparison of DMRG results to exact 
results shows that one often 
comes extremely close to an optimal representation
within the matrix state space (which justifies the usage of the DMRG
algorithm to obtain them), it has been pointed 
out and numerically demonstrated\cite{Duke98,Taka99} that finite-system 
DMRG results can be further improved and better matrix 
product states be produced 
by switching, after convergence is reached, from the 
S$\bullet\bullet$E scheme (with two free sites) 
to an S$\bullet$E scheme and to carry out 
some more sweeps. This point is not pursued further here, it just 
serves to illustrate that finite-system DMRG for all practical 
purposes comes close to an optimal matrix product state, while 
not strictly reaching the optimum.

As the actual decomposition and truncation 
procedure in DMRG and the TEBD simulation algorithm are identical, our 
proposal
is to use the finite-system algorithm to carry out the sequence of 
local time evolutions (instead of, or after, optimizing the ground state), 
thus constructing by Schmidt decomposition and 
truncation new block states best adapted to a state at any given 
point in the time evolution (hence 
adaptive block states) as in the TEBD algorithm, 
while maintaining the computational efficiency
of DMRG. To do this, one needs not only all reduced basis 
transformations, but also the wave function $\ket{\psi}$ in a two-block 
two-site configuration such that the bond that is currently updated 
consists of the two free sites. This implies that $\ket{\psi}$ has to 
be transformed between different configurations. In finite-system 
DMRG such a transformation, which was first implemented 
by White\cite{Whit96} (``state 
prediction'') is routinely used to predict the outcome of large sparse 
matrix diagonalizations, which no longer occur during time 
evolution. Here, it merely serves as a basis transformation. We will 
outline the calculation for shifting the active bond by one site to 
the left.

Starting from 
\begin{equation}
    \ket{\psi} =
    \sum_{m^S_{l-1}} 
    \sum_{\sigma_{l}} 
    \sum_{\sigma_{l+1}} 
    \sum_{m^E_{l+1}} \psi_{m^S_{l-1}\sigma_{l}\sigma_{l+1} m^E_{l+1}} 
    \ket{w^S_{m_{l-1}}}\ket{\sigma_{l}}\ket{\sigma_{l+1}}\ket{w^E_{m_{l+1}}} ,
\end{equation}    
one inserts the identity $\sum_{m_{l}^E} \ket{w^E_{m_{l}}}\bra{w^E_{m_{l}}}$ 
obtained from the Schmidt decomposition (i.e.\ density matrix 
diagonalization) to obtain
\begin{equation}
    \ket{\psi}=
    \sum_{m^S_{l-1}} 
    \sum_{\sigma_{l}} 
    \sum_{m^E_{l}} \psi_{m^S_{l-1}\sigma_{l}m^E_{l}} 
    \ket{w^S_{m_{l-1}}}\ket{\sigma_{l}}\ket{w^E_{m_{l}}} ,
\end{equation}    
where
\begin{equation}
    \psi_{m^S_{l-1}\sigma_{l}m^E_{l}} = \sum_{m^E_{l+1}} \sum_{\sigma_{l+1}}
    \psi_{m^S_{l-1}\sigma_{l}\sigma_{l+1} m^E_{l+1}} 
    A^{l+1}_{m_{l}m_{l+1}}[\sigma_{l+1}] .
\end{equation}    
After inserting in a second step the identity $\sum_{m_{l-2}^S\sigma_{l-1}} 
\ket{w^S_{m_{l-2}}\sigma_{l-1}}\bra{w^S_{m_{l-2}}\sigma_{l-1}}$, 
one ends up with the wave function in the shifted bond representation:
\begin{equation}
    \ket{\psi}=
    \sum_{m^S_{l-2}} 
    \sum_{\sigma_{l-1}} 
    \sum_{\sigma_{l}} 
    \sum_{m^E_{l}} \psi_{m^S_{l-2}\sigma_{l-1}\sigma_{l} m^E_{l}} 
    \ket{w^S_{m_{l-2}}}\ket{\sigma_{l-1}}\ket{\sigma_{l}}\ket{w^E_{m_{l}}} ,
\end{equation}    
where
\begin{equation}
    \psi_{m^S_{l-2}\sigma_{l-1}\sigma_{l} m^E_{l}} = \sum_{m^S_{l-1}}
    \psi_{m^S_{l-1}\sigma_{l}m^E_{l}}  A^{l-1}_{m_{l-2}m_{l-1}}[\sigma_{l-1}] .
\end{equation}

\section{Adaptive time-dependent DMRG}
\label{sec:newtime}
The adaptive time-dependent DMRG algorithm which incorporates the 
TEBD simulation algorithm in the DMRG framework 
is now set up as follows (details 
on the finite-system algorithm can be found in Ref.\ \cite{Whit93}):
\begin{itemize}
\item[0.] Set up a conventional finite-system DMRG algorithm with state 
prediction using the 
Hamiltonian at time $t=0$, $\ham(0)$, to determine the ground state of
some system of length $L$ using effective block Hilbert spaces of 
dimension $M$. At the end of this stage of the algorithm, we have
for blocks of all sizes $l$ reduced orthonormal 
bases spanned by states $\ket{m_{l}}$, which are characterized by good quantum 
numbers. Also, we have all reduced basis transformations, 
corresponding to the matrices $A$.
\item[1.]  For each Trotter time step, use the finite-system DMRG 
algorithm to run one sweep with the following modifications: 

 \begin{itemize}
  \item[i)] For each even bond apply the local time evolution $\hat{U}$ at the 
bond formed by the free sites to $\ket{\psi}$. This is a very fast operation
compared to determining the ground state, which is usually done instead in the
finite-system algorithm.
\item[ii)] As always, perform a DMRG truncation at each step of the 
finite-system algorithm, hence $O(L)$ times.
\item[(iii)] Use White's prediction method to shift the free sites by one.
\end{itemize}
\item[2.] In the reverse direction, apply step (i) to all odd bonds.
\item[3.] As in standard finite-system DMRG evaluate operators when 
desired at the end of some time steps. Note that there is no need to 
generate these operators at all those time steps where no operator 
evaluation is desired, which will, due to the small Trotter time step, 
be the overwhelming majority of steps. 
\end{itemize}

The calculation time of adaptive time-dependent DMRG scales linearly in $L$,
as opposed to the static time-dependent DMRG which does not depend on 
$L$. The diagonalization of the density matrices (Schmidt 
decomposition) scales as $N_{{\rm site}}^3 M^3$; the preparation of 
the local time evolution operator as $N_{{\rm site}}^6$, but this may 
have to be done only rarely e.g.\ for discontinuous changes of 
interaction parameters. Carrying out the local time evolution
scales as $N_{{\rm site}}^4 M^2$; the basis transformation 
scales as $N_{{\rm site}}^2 M^3$. As $M\gg N_{{\rm site}}$ 
typically, the algorithm is of order $O(LN_{{\rm site}}^3 M^3)$ at 
each time step.

\section{Case study: time-dependent Bose-Hubbard model}
\label{sec:bose}
In this section we present some results of calculations on the Bose-Hubbard 
Hamiltonian introduced in section \ref{sec:oldtime} 
which have been carried out, using modest computational resources and an 
unoptimized code (this concerns in particular the operations on 
complex matrices and vectors). In the following, Trotter time steps 
down to $\delta t = 5 \times 10^{-4}$ in units of $\hbar/J$ were chosen.
It is also important to note that in contrast to the DMRG calculations
shown earlier for conventional time-dependent DMRG up to $N_{{\rm site}}=14$ states per 
site were used as a local site basis for all calculations in this 
Section.

\begin{figure}
\centering\epsfig{file=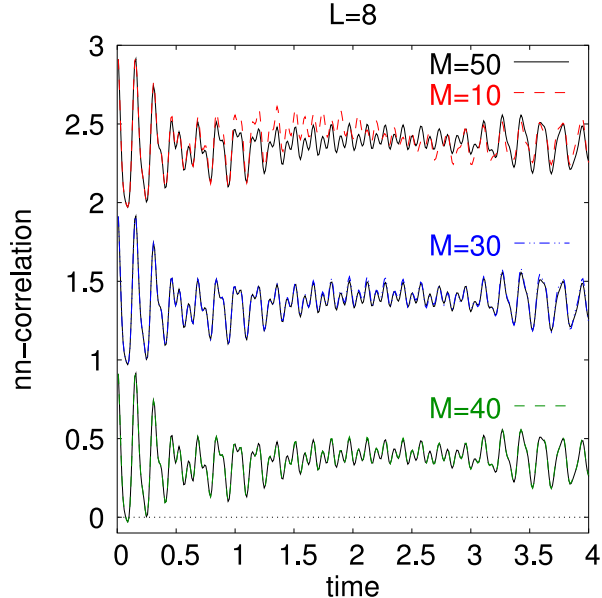,width=0.6\linewidth}
\caption{Time evolution of the real part of nearest-neighbor correlations in a 
Bose-Hubbard model with instantaneous change of interaction strength
using the adaptive time-dependent DMRG. The different curves for different $M$
are shifted (parameters as in section \ref{sec:oldtime}).}
\label{fig:localconvM}
\end{figure}

Comparing the results of the adaptive time-dependent DMRG for the Bose-Hubbard
model with the parameters chosen as in section \ref{sec:oldtime} with the 
static time-dependent DMRG we find that the convergence in $M$ is much faster,
for the nearest neighbor correlations it sets in at about $M=40$ (Fig. \ref{fig:localconvM})  compared to $M=100$ for the static method 
(Fig. \ref{fig:crank3}).

\begin{figure}
\centering\epsfig{file=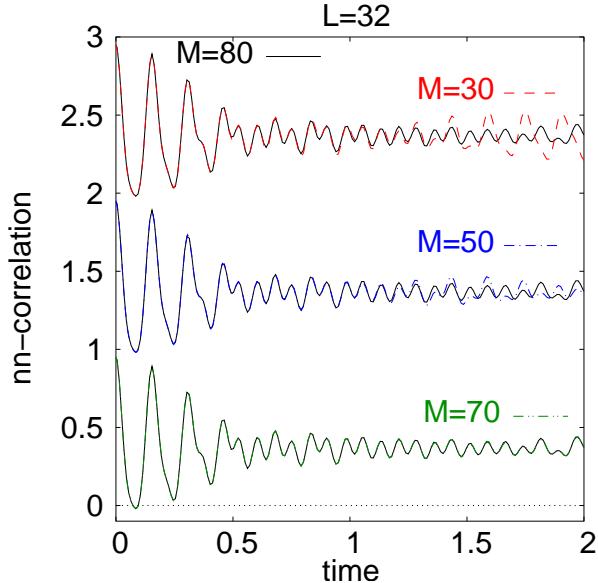,width=0.6\linewidth}
\caption{Time evolution of the real part of nearest-neighbor correlations in a 
Bose-Hubard model with instantaneous change of interaction strength
using the adaptive time-dependent DMRG but for a larger system $L=32$ 
with $N=32$ bosons. The 
different curves for different $M$ are shifted, comparing $M=30,50,70$ 
to $M=80$ respectively.}
\label{fig:localconvML32}
\end{figure}

This faster convergence in $M$ enables us to study larger systems than with 
static time-dependent DMRG (Fig. \ref{fig:localconvML32}). In the 
$L=32$ system considered here, we encountered severe convergence 
problems using static time-dependent DMRG. By contrast, in the new 
approach convergence sets in for $M$ well below 100, which is easily 
accessible numerically. Let us remark that the number $M$ of states which have to be kept does
certainly vary with the exact parameters chosen, 
depending if the state can be approximated well by matrix product states 
of a low dimension. At least in the case studied here, we found that 
this dependency is quite weak. We expect (also from 
studying the time evolution of density matrix spectra) that the model 
dependence of $M$ is roughly similar as in the static case.

\begin{figure}
\centering\epsfig{file=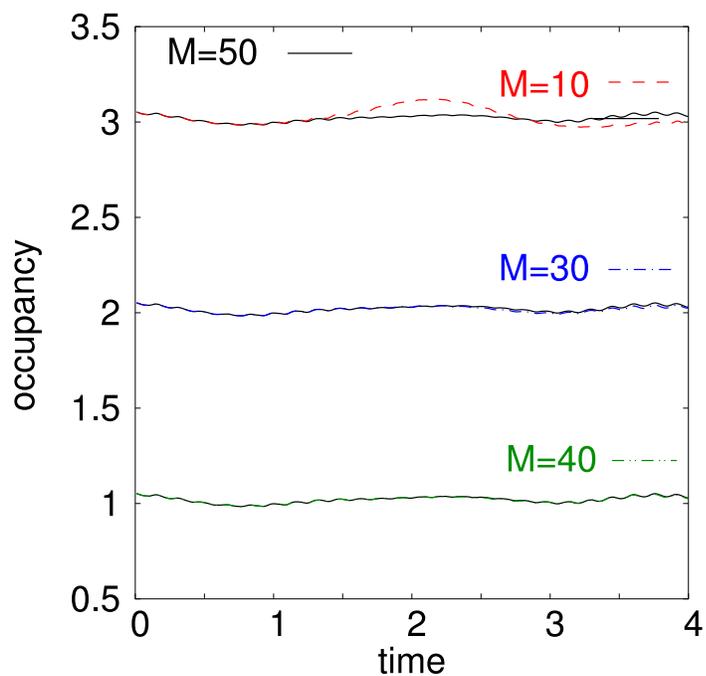,width=0.7\linewidth}
\caption{Time evolution of the occupancy of the second site. Parameters as
  used in section \ref{sec:oldtime} ($L=8$, $N=8$). The different curves for different $M$ are shifted.} 
\label{fig:occupancy}
\end{figure}

\begin{figure}
\centering\epsfig{file=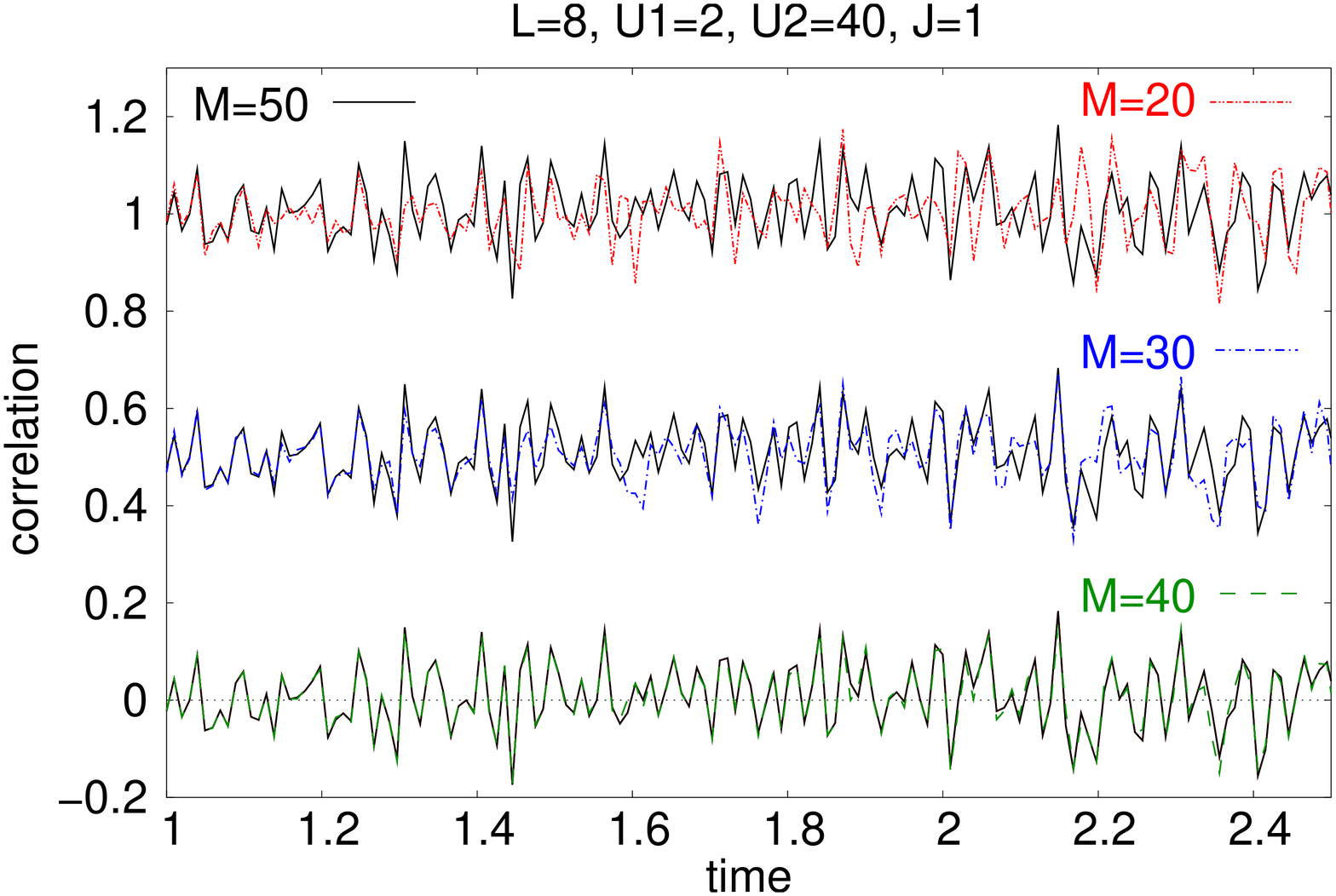,width=\linewidth}
\caption{Time evolution of the real part of the correlation between site 2 and
  7. Parameters as used in section \ref{sec:oldtime} with $N=8$ 
  particles. The different curves for 
  different $M$ are shifted. Note that the plot starts at $t=1$ (parameters
were changed at $t=0$).}
\label{fig:farcorr}
\end{figure}

Similar observations are made both for local occupancy (a simpler 
quantity than nearest-neighbor correlations) and longer-ranged 
correlations (where we expect less precision). Moving back to the 
parameter set of section \ref{sec:oldtime}, we find as expected 
that the result for the local occupancy (Fig. \ref{fig:occupancy}) 
is converged 
for the same $M$ leading to convergence in the nearest-neighbor 
correlations. 
In contrast, if we consider the correlation $\langle b^\dagger b \rangle$ 
between sites further apart from each other the numerical results converge
more slowly under an increase of $M$ than the almost local quantities. 
This can be seen in Fig. \ref{fig:farcorr} where the results for $M=40$ and $M=50$ still differ a
bit for times larger than $t\approx 2 \hbar/J$.

\begin{figure}
\centering\epsfig{file=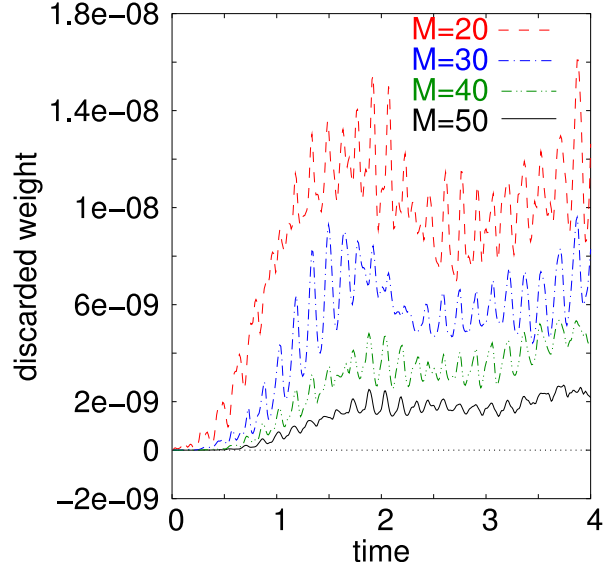,width=0.6\linewidth}
\caption{Discarded weight for different values of $M$. Parameters chosen as in
  section \ref{sec:oldtime}.}
\label{fig:discweight}
\end{figure}

\begin{figure}
\centering\epsfig{file=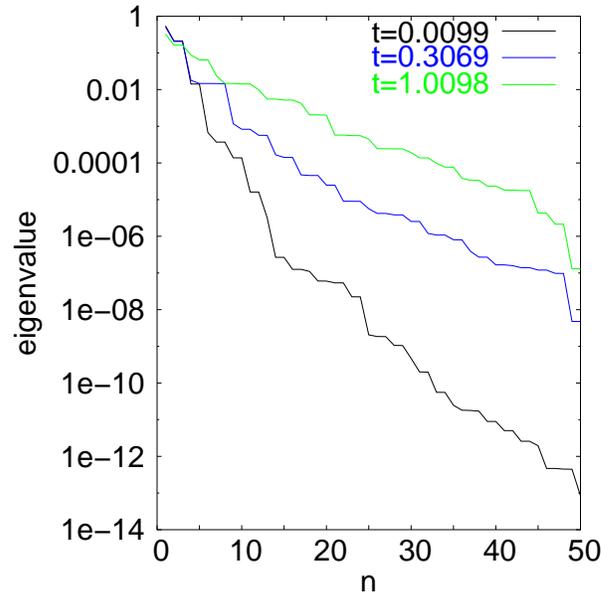,width=0.6\linewidth}
\caption{Eigenvalue spectrum of the left reduced density matrix at different times
for a symmetric S/E decomposition. Parameters chosen as in
  section \ref{sec:oldtime}, $M=50$ states retained.}
\label{fig:spectrum}
\end{figure}

The controlling feature of DMRG is the density matrix formed at each 
DMRG step -- the decay of the density-matrix eigenvalue spectrum
and the truncated weight (i.e.\ the sum of all eigenvalues whose 
eigenvectors are not retained in the block bases) control its 
precision. In the discarded weight for the Bose-Hubbard model 
of section \ref{sec:oldtime} shown in Fig. \ref{fig:discweight},
we can observe that the discarded weight shrinks drastically, 
going from $M=20$ to $M=50$. This supports the idea 
that the system shows a fast 
convergence in $M$. Even more importantly, the discarded weight grows 
in time, as the state that was originally a ground state at $t<0$ decays into 
a superposition of many eigenstates of the system at $t>0$. However, 
in particular for larger $M$, it stays remarkably small throughout the 
simulation, indicating that adaptive time-dependent DMRG tracks the 
time-evolving state with high precision. Moving to the
detailed spectrum of the density matrix (shown in Fig. \ref{fig:spectrum} for
the left density matrix when the chain is symmetrically decomposed 
into S and E), the corresponding distribution of the eigenvalues can be seen 
to be approximately exponential. In agreement with the increasing 
truncation error, one also observes that the decay becomes less steep 
as time grows. Yet, we still find a comparatively fast decay of the 
eigenvalue spectrum at all times, necessary to ensure the applicability 
of TEBD and adaptive time-dependent DMRG respectively. 
  

Note for all results shown that the unusually large number of states per site 
($N_{{\rm site}}=14$) which would not occur in Hubbard or Heisenberg models 
could there be translated directly into longer chains or 
larger state spaces (larger $M$) for the same 
computational effort, given that the algorithm is $O(LN_{{\rm site}}^3 M^3)$.
In that sense, we have been discussing an algorithmically hard case, 
but in fermionic models DMRG experience tells us that $M$ has to be taken 
much larger in fermionic systems. For the fermionic Hubbard model, with 
$N_{{\rm site}}=4$, more than $M=300$ is feasible with the unoptimized code, 
and much higher $M$ values would be possible 
if optimizations were carried out. This should be enough to 
have quantitatively reliable time-evolutions for fermionic chains, 
while of course not reaching the extreme precision one is used to in 
DMRG for the static case. As the algorithmic cost is dominated by 
$(N_{{\rm site}}M)^3$, the product $N_{{\rm site}}M$ is an 
important quantity to look at: while current TEBD implementations 
range at 100 or less, adaptive time-dependent DMRG using good quantum 
numbers runs at the order of 1000 (and more).

\begin{figure}
\centering\epsfig{file=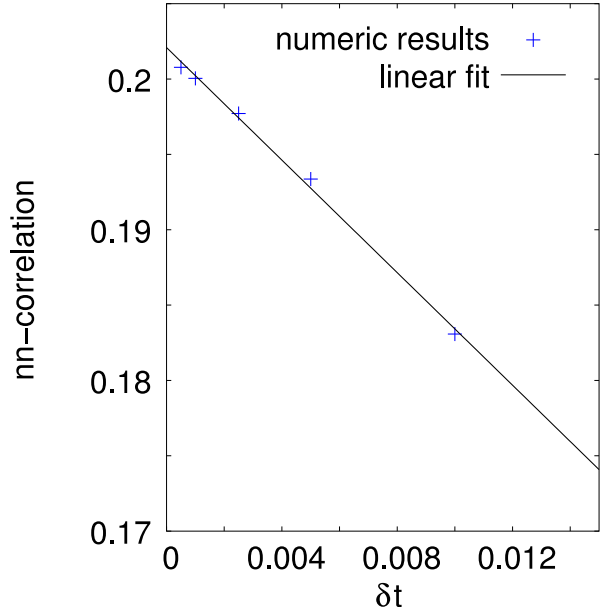,width=0.6\linewidth}
\caption{Convergence in the Trotter time of the real part of the nearest-neighbor correlations
  between site 2 and 3 in a 
Bose-Hubbard model with instantaneous change with the parameters chosen as in
  section \ref{sec:oldtime} at a fixed time.}
\label{fig:epsconv}
\end{figure}

Let us conclude this section by pointing out that at least one 
improvement can be incorporated almost trivially into this most simple 
version of adaptive time-dependent DMRG. Since we have used a first-order 
Trotter decomposition, we expect that for fixed $M$
results of measurements at a fixed time converge linearly with respect 
to the time step $\delta t$ chosen, 
as the error per time step scales as $\delta t^2$, but the 
number of time steps needed to reach the fixed time grows as $\delta t^{-1}$.
In other words, the Trotter error is inversely proportional to the 
calculation time spent. This can indeed be observed in results such as 
presented in Fig. \ref{fig:epsconv}.

It is very easily and at hardly any algorithmic cost that a second order 
Trotter decomposition can be implemented, leading to errors of order
$\delta t^2$. The second order Trotter 
decomposition reads\cite{Suzu76}
\begin{equation}
    e^{-\imag \hat{H}\delta t} =
    e^{-\imag \hat{H}_{odd}\delta t/2}e^{-\imag \hat{H}_{even}\delta 
    t}e^{-\imag \hat{H}_{odd}\delta t/2} ,
\label{eq:secondtrotter}
\end{equation}
where we have grouped all local Hamiltonians on odd and even bonds into
$\hat{H}_{odd}$ and $\hat{H}_{even}$ respectively.
At first sight this seems to indicate that at each Trotter time step 
three (instead of two) moves (``zips'') through the chain have to be carried out.
However, in many applications at the end of most time steps, 
the Hamiltonian does not change, 
such that for almost all time steps, we can 
contract the second $e^{-\imag \hat{H}_{odd}\delta t/2}$ from the 
previous and the first $e^{-\imag \hat{H}_{odd}\delta t/2}$ from the 
current time step to a standard $e^{-\imag \hat{H}_{odd}\delta t}$ 
time step. Hence, we incur almost no algorithmic cost. 
This is also standard practice in Quantum Monte Carlo \cite{Assa91}; 
following QMC, second order Trotter evolution 
is set up as follows:
\begin{enumerate}
    \item Start with a half-time step $e^{-\imag \hat{H}_{odd}\delta t/2}$.
    \item Carry out successive time steps $e^{-\imag \hat{H}_{even}\delta 
    t}$ and $e^{-\imag \hat{H}_{odd}\delta 
    t}$.
    \item At measuring times, measure expectation values 
    after a $e^{-\imag \hat{H}_{odd}\delta 
    t}$ time step, and again after a time step $e^{-\imag 
    \hat{H}_{even}\delta t}$,
    and form the average of the two values as the outcome of the 
    measurement.
    \item At times when the Hamiltonian changes, do not contract two 
    half-time steps into one time step.
\end{enumerate}
In this way, additional algorithmic cost is only incurred at the (in 
many applications rare) times when the Hamiltonian changes 
while strongly reducing the Trotter decomposition error.
Even more precise, but now at an algorithmic cost of factor 5 over 
the first or second-order decompositions, would be the usage of
fourth-order Trotter decompositions (leading to 15 zips through the 
chain per time step, 
of which 5, however, can typically be eliminated)\cite{Yosh90,Krec98}.

\section{Conclusion}
\label{sec:conclusion}
The TEBD algorithm for the simulation of slightly entangled quantum 
systems, such as quantum spin chains and other one-dimensional quantum 
systems, was originally developed in order to establish a link between the 
computational potential of quantum systems and their degree of 
entanglement, and serves therefore as a good example of how concepts and 
tools from quantum information science can influence other areas of 
research, in this case quantum many-body physics. 

While exporting ideas from one field of knowledge to another may appear as 
an exciting and often fruitful enterprise, differences in language and 
background between researchers in so far separated fields can also often 
become a serious obstacle to the proper propagation and full assimilation 
of such ideas. In this paper we have translated the TEBD algorithm into 
the language of matrix product states. This language is a natural choice 
to express the DMRG algorithm -- which, for over a decade, has dominated 
the simulation of one-dimensional quantum many-body systems. In this way, 
we have made the TEBD algorithm fully accessible to the DMRG community. On 
the other hand, this translation has made evident that the TEBD and the 
DMRG algorithms have a number of common features, a fact that can be 
exploited.

We have demonstrated that a very straightforward modification of existing 
finite-system DMRG codes to incorporate the TEBD leads to a new adaptive 
time-dependent DMRG algorithm. Even without attempting to reach the 
computationally most efficient incorporation of the TEBD
algorithm into DMRG implementations, the resulting code seems to perform 
systematically better than static 
time-dependent DMRG codes at very reasonable numerical cost, 
converging for much smaller state spaces, as they change in time to 
track the actual state of the system. 
On the other hand, while it presents no new conceptual idea, the new code 
is also significantly more efficient than existing embodiments of the 
TEBD, for instance thanks to the way DMRG handles good quantum numbers.
While we have considered bosons as an example, as in standard DMRG fermionic 
and spin systems present no additional difficulties. 
Various simple further improvements are feasible, and 
we think that adaptive time-dependent DMRG can be applied not only to 
problems with explicitly time-dependent Hamiltonians, but also to 
problems where the quantum state changes strongly in time, such as in 
systems where the initial quantum state is far from equilibrium. The 
method should thus also be of great use in the fields of transport and 
driven dissipative quantum systems.

{\em Acknowledgments.} US and GV would wish to thank the Institute of 
Theoretical Physics at the University of Innsbruck, where this work 
was initiated, for its hospitality. CK is supported by the 
Studienstiftung des Deutschen Volkes and DFG grant DE 730/3-1. The work in
Innsbruck is 
supported by EU networks and the Institute for Quantum Information. 
GV acknowledges support from the US National Science 
Foundation under Grant No. EIA-0086038. US acknowledges support by The 
Young Academy at the Berlin-Brandenburg Academy of Sciences and the 
Leopoldina Society. We also thank
Peter Zoller, Dieter Jaksch, Hans Briegel, Willi Zwerger, Ignacio 
Cirac, Juanjo Garcia-Ripoll, Miguel Cazalilla, Brad Marston, Jan von Delft and 
Matthias Troyer for discussions.

{\em Note added in proof.} After submission of this work, we became 
aware of closely related work by White and Feiguin\cite{Whit04}.

\end{document}